       \documentstyle[]{l-aa-dipastro}
       \input psfig.tex
         
	\newcommand{\Zsun}{\mbox{${\rm Z_{\odot}}$}}

        \newcommand{\Hbeta}{\mbox{${\rm H_{\beta}}$}}
        \newcommand{\MFe}{\mbox{${\rm \langle Fe \rangle}$}}
        
        \newcommand{\alfa}{\mbox{$\alpha$-elements}}

        \def\M12{${\rm M_{L,12}} $}
        \def\Mg2{${\rm Mg_{2}}$}

        \newcommand{\dmg}{\mbox{${\rm \delta Mg_{2}}$}}
	\newcommand{\dhb}{\mbox{${\rm \delta H_{\beta}}$}}
	\newcommand{\dfem}{\mbox{${\rm \delta \langle Fe \rangle}$}}
	
	\newcommand{\Dlogt}{\mbox{$\rm \Delta \log(t)$}}

\def\smallskip{\vskip 8pt}
\def\littleskip{\vskip 6pt}

\begin{document}

  \thesaurus{}

   \title{Ages and Metallicities in Elliptical Galaxies from the
           \Hbeta, \MFe, and \Mg2\ Diagnostics}

  \author{ Rosaria Tantalo$^1$, Cesare Chiosi$^{2,1}$, Alessandro Bressan$^3$}

   \institute{
   $^1$ Department of Astronomy, Vicolo dell' Osservatorio 5, 35122 Padua, 
            Italy\\
   $^2$ European Southern Observatory, K-Schwarzschild-strasse 2, D-85748, 
             Garching bei Muenchen, Germany\\
   $^3$ Astronomical Observatory, Vicolo dell' Osservatorio 5, 35122 Padua, 
            Italy}

   \offprints{C. Chiosi }

   \date{Received: November 1997.  Accepted: }

    \maketitle

    \markboth{}{}

\begin{abstract}
Systematic variations in  the line strength indices \Hbeta, \Mg2, and \MFe\
are observed across  elliptical galaxies and  limited to the central regions
passing from one object to another. Furthermore, since the gradients in \Mg2\
and \MFe\ have often different slopes arguments are given for an enhancement
of Mg (\alfa\ in general) with respect to Fe toward the center of these
galaxies. Finally, the inferred degree of enhancement seems to increase
passing from dwarfs to massive ellipticals.
 
In this study we have  investigated the ability of the $\rm H_{\beta}$, \Mg2\
and \MFe\ diagnostics to assess the metallicity, [Mg/Fe]  ratios, and ages of
elliptical galaxies.

To this aim, first we derive  basic calibrations for the variations $\rm
\delta H_{\beta}$, $\rm  \delta Mg_2$ and \dfem\ as a function of variation in
 age \Dlogt, metallicity $\rm \Delta \log(Z/Z_{\odot})$,  and $\rm \Delta
[Mg/Fe]$.

Second, examining the gradients observed in a small sample of galaxies,  we
analyze how the difference $\rm \delta H_{\beta}$, \dmg, and $\rm \delta
\langle Fe \rangle$ between the external and central values of each index
translates into $\rm \Delta [Mg/Fe]$,  $\rm \Delta \log(Z/Z_{\odot})$, and
$\rm  \Delta \log(t)$.  We find that out of six galaxies under examination,
four have the nuclear region more metal-rich, more enhanced in \alfa, and
younger (i.e. containing a significant fraction of stars of relatively young
age) than the external regions. In contrast the remaining two galaxies have
the nuclear region more metal-rich, more enhanced in \alfa\ but marginally
older than the external zones.

Third,  we explore the variation from galaxy to galaxy of the nuclear values
of \Hbeta, \Mg2, and \MFe\ limited to a sub-sample of the Gonz\'ales (1993)
list. The differences \dhb, \dmg, and $\rm \delta \langle Fe \rangle$ are
converted into the differences \Dlogt, $\rm \Delta \log(Z/Z_{\odot})$, and
$\rm \Delta [Mg/Fe]$. Various correlations among the age, metallicity, and
enhancement variations are explored. In particular we thoroughly examine the
relationships $\rm \Delta\log(t)-M_V$, $\rm \Delta \log(Z/Z_{\odot})-M_V$, and
$\rm \Delta [Mg/Fe]-M_V$. It is found that a sort of age limit is likely to
exist in the  $\rm \Delta\log(t)-M_V$ plane, traced by galaxies with mild or
no sign of rejuvenation. In these objects, the duration of the star forming
activity is likely to have increased at decreasing galactic mass. Limited to
these galaxies, the mass-metallicity sequence implied by the color-magnitude 
relation  is recovered, likewise for the $\alpha$-enhancement-luminosity
relation suggested by the  gradients in \Mg2\ and \MFe. For the remaining
galaxies the situation is more intrigued: sporadic episodes of star formations
are likely to have  occurred scattering the galaxies in the space of age
(\Hbeta), metallicity, and [Mg/Fe].

The results are discussed in regard to  predictions from the merger and
isolation models of galaxy formation and evolution highlighting points of
difficulty with each scheme. Finally, the suggestion is advanced that models
with an IMF that at the early epochs favors higher mass stars in massive
ellipticals galaxies, and lower mass stars in low-mass ellipticals,  might be
able to alleviate some of the difficulties encountered by  the standard
SN-driven galactic wind model and lead to  a coherent interpretation of the
data.

\end{abstract}

\section{Introduction}

Elliptical galaxies exhibit systematic variations in  the line strength indices 
\Hbeta, \Mg2, and \MFe\ either going from the center to the external regions of 
a galaxy or limited to the central regions passing from one galaxy to another.
(cf. Gonz\'ales 1993; Carollo \& Danziger 1994a,b; Carollo et al. 1993; Davies 
et al. 1993; Fisher et al. 1995, 1996). Furthermore, since the gradients in \Mg2\ 
and \MFe\ have different slopes arguments are given for an enhancement of Mg 
(\alfa\ in general) with respect to Fe toward the center of these galaxies. 
Finally, the inferred degree of enhancement seems to increase passing from dwarfs to
massive ellipticals (see Faber et al. 1992, Worthey et al. 1994 and Matteucci
1997 for recent reviews of all these subjects and exhaustive referencing).

All this suggests the occurrence of changes in some fundamental properties
of  the constituent stellar populations. The understanding of this 
problem is that either the age or the metallicity (the abundance pattern, 
in general) or both are likely to be responsible for the above
variations.

Owing to the primary importance of ranking galaxies  (or regions of these) 
as function of the age, metallicity and abundance ratios in regard to the more 
general subject of galaxy formation and evolution, in this paper we present a 
new attempt aimed at unraveling the information contained in the indices \Hbeta,
\Mg2, and \MFe\ for a sample of elliptical galaxies.

The above indices have complicated dependences on the temperature, gravity,
and chemical abundances of the emitting stars, cf. the popular empirical 
calibrations by Worthey (1992), Worthey et al. (1994) and Borges et al. (1995). 
When these calibrations are applied to derive the mean value of the indices 
in question for single stellar populations (SSP) as a function of the age and 
chemical abundances, the results suffer from a certain degree of degeneracy in 
the sense that all the indices vary in the same way at varying age and chemical 
composition, the well known {\it age-metallicity degeneracy} (cf. Worthey 1994). 
Furthermore, when the last step is undertaken, i.e. the above indices are derived 
for complex assemblies of stars of any age and composition, interpreting the 
observational data is even more intrigued because the indices are found to depend 
on the underlying relative distribution of  stars per metallicity bin, i.e. the 
so-called partition function $N(Z)$ introduced by Tantalo et al. (1998) in their 
analysis of this problem. This means that hints on the past history of star formation 
are needed to solve the problem.

Despite these intrinsic difficulties \Hbeta, \Mg2, and \MFe\ respond in different 
fashion to different parameters: \Hbeta\ mainly correlates with the age because it
is most sensitive to the light emitted by stars at the turnoff, whereas \Mg2\ and 
\MFe\ are more related to stars along the RGB and therefore are more sensitive to 
metallicity and abundance ratios (cf. Worthey 1994). Therefore, an attempt to infer 
age, metallicity, and abundance ratios from these line strength indices is possible.
 
The plan of the paper is as follows. In Sect. 2 we briefly recall the sources of 
data we have adopted in this study. In Sect. 3 we derive new basic relationships 
providing the variation of \Hbeta, \Mg2, and \MFe\ in SSPs (from which eventually
the galactic indices are built up) as a function of age $t$, total metallicity
Z, and [Mg/Fe], and in Sect. 4 we present the $\Delta$-method. In Sect. 5, we 
apply our diagnostics to  study the gradients in \Mg2 and \MFe\ across elliptical 
galaxies limited to a small number of objects for which the data were available. 
In Sect. 6, we study the variations of nuclear values of \Hbeta, \Mg2\ and \MFe\ from
galaxy to galaxy of the Gonz\'ales (1993) sample and translate these variations
into differences in age, metallicity, and enhancement of \alfa. In Sect. 7, we 
thoroughly examine the galaxies in the space of $\rm \Delta log(t)$, 
$\rm \Delta log(Z/\Zsun)$, and $\rm \Delta [Mg/Fe]$, where $\rm \Delta$ stands for 
the difference between the values of the three quantities for each individual object 
and their mean values in the sample. In Sect. 8 we attempt a preliminary ranking of 
the galaxies as a function of their absolute age, metallicity, and [Mg/Fe]. In Sect. 9 
we discuss the results of the analysis in regard to the present-day understanding of 
the mechanism of formation and evolution of elliptical galaxies: i.e. isolation or 
merger. In Sect. 10 we compare the results with the predictions from the SN-driven 
galactic wind both at constant and variable IMF. Finally, some concluding remarks 
are drawn in Sect. 11.

\section{ The observational data}
There are several sources of data that are useful to our purposes:

\begin{itemize}
\item{ The fully corrected indices of Gonz\'ales (1993) for two different galaxy 
coverage. The central region within Re/8 (Re is the effective radius), 
the wider area within Re/2, and finally the {\it nuclear region} 
within $2\times 4.1$ arcsec. In the same source there are high quality 
gradient data for some 20 galaxies. The Gonz\'ales (1993) list of nuclear
data is used to study the variations in the indices from galaxy to galaxy.}
 
\item{ The small sample of ellipticals by Carollo \& Danziger (1994a,b) 
containing detailed data for gradients in \Mg2\ and \MFe. This sample is utilized 
to illustrate the capability of the method when applied to individual galaxies,
and to perform a preliminary analysis of the gradient in age, metallicity and ratio [Mg/Fe]
in individual objects.}

\item{ The samples by Davies et al. (1993) and Fisher et al. (1995, 1996) containing 
data on 13 and 7 elliptical respectively, and finally galaxies by Vazdekis (1997,
private communication). The results for these data sets are not
shown here for the sake of brevity. They have been  separately used to check
whether results obtained with the other sets were also confirmed by these
ones. }
\end{itemize}

No attempt is made to put all these data sources together, because 
we feel that manipulating the data to render all sources mutually
consistent (it is worth  recalling  that each sample differs from the others 
in many observational details: galaxy coverage, reduction technique, calibrations, 
etc..) would introduce spurious effects difficult to handle. For these 
reasons we prefer to use smaller albeit homogeneous sets of data. It is 
beyond the aims of this study to perform a systematic analysis of all data in
literature. We anticipate that the results obtained from using separately each 
sub-group of data are in mutual agreement.

\section{Calibrations containing [Mg/Fe]}
Many studies have emphasized that line strength indices depend not only on the
stellar parameters $T_{eff}$, gravity, and [Fe/H] as in the classical 
calibration by Worthey (1992) and Worthey et al. (1994), but also on the 
ratios of chemical abundances such as [Mg/Fe] in particular
(Barbuy 1994, Idiart et al. 1995, Weiss et al. 1995, Borges et al. 1995,
de Freitas \& Pacheco 1997, Idiart \& de Freitas Pacheco 1997).
Needless to say that for the purposes of this study we have to adopt 
calibrations in which the ratio [Mg/Fe] is taken into account.

We start pointing out that in presence of a certain degree of enhancement in
$\alpha$-elements one has to suitably modify relationship between the total
metallicity Z and the iron content [Fe/H]. Using the pattern of abundances
by Anders \& Grevesse (1989), Grevesse (1991) and Grevesse \& Noels (1993), we
find the general relation

\begin{equation}
\rm \left[\frac{Fe}{H} \right] = \log{\left(\frac{Z}{Z_{\odot}}\right)} -
 \log{\left(\frac{X}{X_{\odot}}\right)} -
0.8\left[\frac{\alpha}{Fe}\right] - 0.05\left[\frac{\alpha}{Fe} \right]^{2}
\label{feh}
\end{equation}

\noindent 
where the term $\rm [\alpha/Fe]$ stands for all $\alpha$-elements lumped together.
 
The recent empirical calibration by Borges et al. (1995) for the \Mg2\ index
includes the effect of different [Mg/Fe] ratios

\begin{displaymath}
\rm  {\ln}{Mg_{2}} = -9.037 + 5.795 \frac{5040}{T_{eff}} + 0.398 \log{g}
+ ~~~~~~~~
\end{displaymath}
\begin{equation}
\rm  ~~~~~~~~0.389 \left[ \frac{Fe}{H} \right] - 0.16 \left[ \frac{Fe}{H}
\right]^{2} + 0.981 \left[ \frac{Mg}{Fe} \right] 
\label{mg2}
\end{equation}

\noindent
which holds for effective temperatures and gravities in the ranges $ 3800  <
T_{eff} < 6500$ K and $ 0.7 < \log{g} < 4.5$.

No corresponding calibrations for \Hbeta\ and \MFe\ are available in Borges et 
al. (1995). Some insight on the dependence of \Hbeta\ and \MFe\ on the abundance 
of Mg and Fe can be derived from Tripicco \& Bell (1995). In that paper, one can 
see that \MFe\ is virtually unaffected by the abundance of Mg (it does of course 
on Fe), whereas \Hbeta\ has a more complicated behaviour. It is virtually unaffected
by Mg in cool dwarf and turn-off  stars, and slightly anti-correlated to Mg in 
cool giants. This index does not respond to variations in Fe in cool dwarf stars,
whereas it slightly decreases and increases with Fe in turn-off and cool giant stars, 
respectively.    

However, as the results by Tripicco \& Bell (1995) cannot be easily implemented in our
code to calculate the indices for SSPs (because of the insufficient coverage of 
effective temperature and gravity), we decided not to use the Tripicco \& Bell (1995)
data, and to   adopt the calibrations  by Worthey et al. (1994) with no dependence on 
[Mg/Fe]. With this choice, we limit the effect of [Mg/Fe] on \Hbeta\ and \MFe\ to the 
zero-order evaluation via the different relation between Z and [Fe/H] of the 
$\rm [\alpha/Fe] \neq 0$ case (cf. Tantalo et al. 1998 for more details), and postpone 
to a future study the evaluation of the full effect of [Mg/Fe] on the indices in question 
when the desidered calibrations will be available. 

With the aid of the relations above for [Fe/H], \Mg2\ and implicitly \Hbeta\ and \MFe\ 
are used to generate new SSPs in which not only the chemical abundances are enhanced with 
respect to the solar value but also the effect of this on the line strength indices is taken 
into account in a self-consistent manner.

The procedure is as follows: assumed the total metallicity Z and enhancement ratio [Mg/Fe] as 
input parameters for the chemical abundances, we derive from eq. (\ref{feh}) the 
corresponding value of [Fe/H] to be inserted in the calibrations of Borges et al. (1995) for 
\Mg2 (eq.~\ref{mg2}), and Worthey (1992) and Worthey et al. (1994) for \MFe\ and \Hbeta. 
The helium content Y associated to each value of Z is the same as in the standard SSP of 
Bertelli et al. (1994). Table~1 lists a summary of the parameters characterizing each SSP 
and their \Mg2, \MFe\ and \Hbeta\ for four values of age (15, 10, 5, and 1 Gyr).

It it worth clarifying that these SSPs are calculated with the stellar models of the 
Padua library (Bertelli et al. 1994) in which $\rm [\alpha/Fe]=0$. Although full 
consistency would require that the stellar models in usage had the same 
abundance ratios as the SSPs above, neglecting this has very marginal 
consequences as models calculated with the new opacity tables of
the Livermore group (Iglesias \& Rogers 1993), in which the enhancement 
of the \alfa\ is taken into account, and the re-definition of [$\alpha$/H] at fixed 
Z  are virtually indistinguishable in the CMD from the old ones (Bressan et al. 1997).
The marginal variation in the lifetime of the core H-burning phase can be
ignored for all practical purposes.

\begin{table}
\begin{center}
\caption{New SSPs with different metallicity and [Mg/Fe].}
\scriptsize
\begin{tabular}{c| l| r| c| c| c| c}
\hline
\hline
 & & & & & & \\
\multicolumn{1}{c|}{Age(Gyr)} &
\multicolumn{1}{c|}{Z} &
\multicolumn{1}{c|}{[Mg/Fe]} &
\multicolumn{1}{c|}{$\rm [Fe/H]$} &
\multicolumn{1}{c|}{\Mg2} &
\multicolumn{1}{c|}{\MFe} &
\multicolumn{1}{c}{\Hbeta} \\
\hline
 & & & & & & \\
15 & 0.004 &  +0.3 & --0.952 & 0.134 & 1.714 & 1.687 \\
15 & 0.004 &   0.0 & --0.707 & 0.125 & 1.931 & 1.710 \\
15 & 0.004 & --0.3 & --0.472 & 0.115 & 2.165 & 1.750 \\
15 & 0.02  &  +0.3 & --0.214 & 0.262 & 2.787 & 1.312 \\
15 & 0.02  &   0.0 &  +0.031 & 0.229 & 3.087 & 1.377 \\
15 & 0.02  & --0.3 &  +0.266 & 0.198 & 3.395 & 1.452 \\
15 & 0.05  &  +0.3 &  +0.253 & 0.365 & 3.606 & 1.197 \\
15 & 0.05  &   0.0 &  +0.497 & 0.308 & 3.937 & 1.279 \\
15 & 0.05  & --0.3 &  +0.733 & 0.258 & 4.276 & 1.368 \\
 & & & & & & \\
10 & 0.004 &  +0.3 & --0.952 & 0.126 & 1.609 & 1.884 \\
10 & 0.004 &   0.0 & --0.707 & 0.116 & 1.839 & 1.910 \\
10 & 0.004 & --0.3 & --0.472 & 0.106 & 2.084 & 1.954 \\
10 & 0.02  &  +0.3 & --0.214 & 0.243 & 2.649 & 1.489 \\
10 & 0.02  &   0.0 &  +0.031 & 0.212 & 2.949 & 1.556 \\
10 & 0.02  & --0.3 &  +0.266 & 0.183 & 3.262 & 1.629 \\
10 & 0.05  &  +0.3 &  +0.253 & 0.357 & 3.503 & 1.312 \\
10 & 0.05  &   0.0 &  +0.497 & 0.300 & 3.840 & 1.393 \\
10 & 0.05  & --0.3 &  +0.733 & 0.250 & 4.184 & 1.483 \\
 & & & & & & \\
 5 & 0.004 &  +0.3 & --0.952 & 0.104 & 1.401 & 2.388 \\
 5 & 0.004 &   0.0 & --0.707 & 0.097 & 1.630 & 2.415 \\
 5 & 0.004 & --0.3 & --0.472 & 0.090 & 1.875 & 2.455 \\
 5 & 0.02  &  +0.3 & --0.214 & 0.211 & 2.431 & 1.884 \\
 5 & 0.02  &   0.0 &  +0.031 & 0.182 & 2.734 & 1.945 \\
 5 & 0.02  & --0.3 &  +0.266 & 0.155 & 3.050 & 2.018 \\
 5 & 0.05  &  +0.3 &  +0.253 & 0.308 & 3.235 & 1.648 \\
 5 & 0.05  &   0.0 &  +0.497 & 0.255 & 3.584 & 1.734 \\
 5 & 0.05  & --0.3 &  +0.733 & 0.210 & 3.943 & 1.828 \\
 & & & & & & \\
 1 & 0.004 &  +0.3 & --0.952 & 0.051 & 0.659 & 4.819 \\
 1 & 0.004 &   0.0 & --0.707 & 0.046 & 0.823 & 4.819 \\
 1 & 0.004 & --0.3 & --0.472 & 0.042 & 1.015 & 4.831 \\
 1 & 0.02  &  +0.3 & --0.214 & 0.101 & 1.562 & 4.018 \\
 1 & 0.02  &   0.0 &  +0.031 & 0.093 & 1.856 & 4.074 \\
 1 & 0.02  & --0.3 &  +0.266 & 0.085 & 2.166 & 4.140 \\
 1 & 0.05  &  +0.3 &  +0.253 & 0.161 & 2.377 & 3.758 \\
 1 & 0.05  &   0.0 &  +0.497 & 0.142 & 2.731 & 3.532 \\
 1 & 0.05  & --0.3 &  +0.733 & 0.127 & 3.100 & 3.631 \\
 & & & & & & \\
\hline
\hline
\end{tabular}
\end{center}
\label{tab4}
\end{table}

\section{The \dhb, \dmg, and \dfem\ method}
In this section we present the tool utilized to transfer variations in
\Hbeta, \Mg2, and \MFe\ into variations in age (t), metallicity (Z), and [Mg/Fe].
To this purpose we calculate the partial derivatives $\partial Mg_{2}$,
$\partial \langle Fe \rangle$ and $\partial H_{\beta}$ with respect to
metallicity, age, and [Mg/Fe] of SSPs
\littleskip

\begin{displaymath}
\frac{\partial Mg_{2}}{\partial \log{Z/Z_{\odot}}} \biggr|_{t,[\frac{Mg}{Fe}]}
                          \,\,\,\,\,\,\,\,\,\,\,
\frac{\partial Mg_{2}}{\partial [Mg/Fe]} \biggr|_{t,Z} \,\,\,\,\,\,\,\,\,\,\,
\frac{\partial Mg_{2}}{\partial \log{t}} \biggr|_{Z,[\frac{Mg}{Fe}]}
\label{dlgmgz}
\end{displaymath}
\noindent

\begin{displaymath}
\frac{\partial \langle Fe \rangle}{\partial \log{Z/Z_{\odot}}} 
             \biggr|_{t,[\frac{Mg}{Fe}]}\,\,\,\,\,\,\,\,\,\,\,
\frac{\partial \langle Fe \rangle}{\partial [Mg/Fe]} \biggr|_{t,Z}
\,\,\,\,\,\,\,\,\,\,\,  \frac{\partial \langle Fe \rangle}{\partial \log{t}}
\biggr|_{Z,[\frac{Mg}{Fe}]}
\label{dlgfez}
\end{displaymath}

\begin{displaymath}
\frac{\partial H_{\beta}}{\partial \log{Z/Z_{\odot}}} 
            \biggr|_{t,[\frac{Mg}{Fe}]}\,\,\,\,\,\,\,\,\,\,\,
\frac{\partial H_{\beta}}{\partial [Mg/Fe]} 
            \biggr|_{t,Z} \,\,\,\,\,\,\,\,\,\,\,
\frac{\partial H_{\beta}}{\partial \log{t}} \biggr|_{Z,[\frac{Mg}{Fe}]}
\label{dhbeta}
\end{displaymath}

\littleskip

\noindent
together with their uncertainties due to the small deviations of the
various quantities over the range of age, Z, and [Mg/Fe] spanned by the 
calibrating SSPs (cf. the entries of Table~1). 

The  mean variation \dmg, \dfem\ and \dhb\ as a function of [Mg/Fe], 
$\rm log(Z/\Zsun)$, and $\rm log(t)$, where $\Zsun=0.016$
and ages are Gyr, are cast as follows

\littleskip
\begin{displaymath}
\rm \delta Mg_{2} = ~0.0994 (\pm 0.0003) \times \Delta \left[ \frac{Mg}{Fe}
\right] 
\end{displaymath}
\begin{displaymath}
\rm ~~~~~~~~~~~ + 0.1660 (\pm 0.0299) \times \Delta \log{\left(
\frac{Z}{\Zsun}\right)} 
\end{displaymath}
\begin{equation}
\rm ~~~~~~~~~~~ + 0.0889 (\pm 0.0047) \times \Delta \log(t)
\label{dmg}
\end{equation}
\littleskip

\begin{displaymath}
\rm \delta \langle Fe \rangle = -0.9806 (\pm 0.0003) \times 
                     \Delta \left[ \frac{Mg}{Fe} \right]  
\end{displaymath}
\begin{displaymath}
\rm ~~~~~~~~~~~ + 1.8743 (\pm 0.2637) \times \Delta \log{\left( \frac{Z}{\Zsun}
\right)} 
\end{displaymath}
\begin{equation}
\rm ~~~~~~~~~~~ + 0.7021 (\pm 0.0079) \times \Delta \log(t)
\label{dfe}
\end{equation}
\littleskip

\begin{displaymath}
\rm \delta H_{\beta} = -0.2097 (\pm 0.0016) \times \Delta \left[ \frac{Mg}{Fe}
\right] 
\end{displaymath}
\begin{displaymath}
\rm ~~~~~~~~~~~ - 0.4837 (\pm 0.0937) \times \Delta \log{\left( \frac{Z}{\Zsun}
\right)} 
\end{displaymath}
\begin{equation}
\rm ~~~~~~~~~~~ - 1.2070 (\pm 0.0084) \times \Delta \log(t)
\label{dhb}
\end{equation}

\littleskip

\noindent
These equations are referred to as the $\Delta$-method.

\begin{figure}
\psfig{file=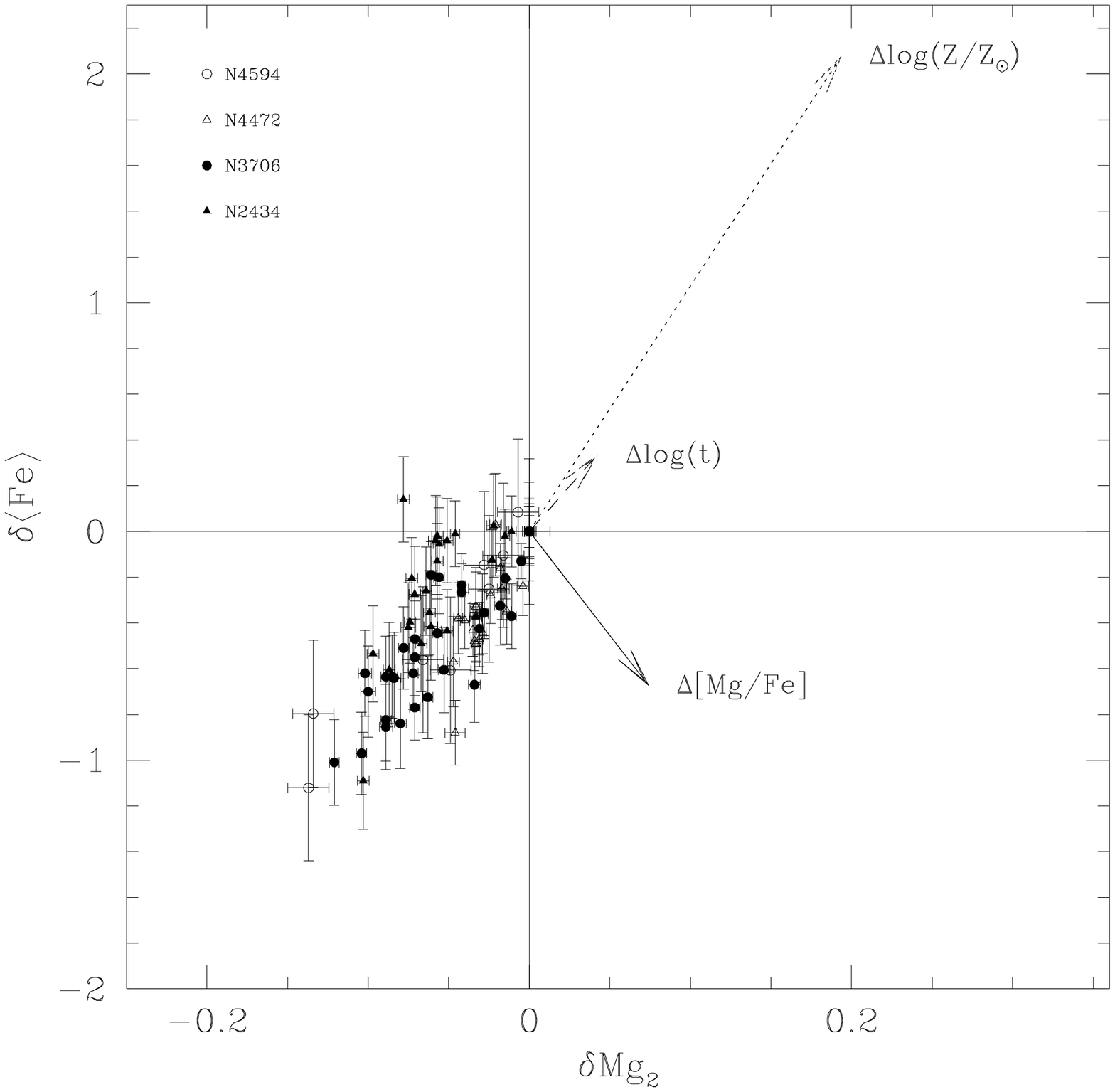,height=9.0truecm,width=8.5truecm}
\caption{The \dfem\ versus $\rm \delta Mg_{2}$ relation across individual galaxies. 
The three arrows centered on (0,0) indicate the {\it age, metallicity, and 
enhancement vectors} as indicated. Along the age vector, the age goes from 5 to 
15 Gyr, along the metallicity vector Z ranges from 0.004 to 0.05, and along the 
enhancement vector [Mg/Fe] goes from $-0.3$ to 0.4 dex. The data are from
Vazdekis (1997, private communication) for NGC~4594, from Davies et al. (1993) 
for NGC~4472, and from Carollo \& Danziger (1994a) for NGC~2434 and NGC~3706.}
\label{dlog}
\end{figure}

To  visually show the size of \dfem\ and $\rm \delta Mg_{2}$ at
varying [Mg/Fe], Z, and log(t), we calculate the {\it age, metallicity,  and
enhancement  vectors} shown in  Fig.~\ref{dlog} for fixed variations of age, 
metallicity, and [Mg/Fe]: the age goes from 5 to 15 Gyr, the metallicity from
0.004 to 0.05, and the [Mg/Fe] ratio from $-0.3$ to $0.4$  dex (the length of
the vectors is $\rm \Delta \log(Z) \sim 1.1$, $\rm \Delta  [Mg/Fe] \sim 0.7$
and $\rm \Delta \log(t) \sim 0.47$). The vectors are centered on (0,0), i.e.
null variation.

There are two important  points to remark:
\begin{itemize}
\item The well known age-metallicity degeneracy gives that the age and
metallicity vectors run very close each other.

\item The enhancement vector is almost orthogonal to the other two, which
allows us to separate its effects from the combined ones of age and
metallicity.
\end{itemize}

Finally, concluding this section we like to recast the system of equations (3)
through (5) so that the theoretical [Mg/Fe], $log(Z/\Zsun)$, and
$log(t)$ are expressed as a function of the observational mean 
variations \dmg, \dfem\ and \dhb 

\begin{displaymath}
\rm \Delta [Mg/Fe] = 5.7410 \times \delta Mg_{2} - 0.4699 \times 
                       \delta \langle Fe \rangle 
\end{displaymath}
\begin{equation}
\rm ~~~~~~~~~ + 0.1495 \times \delta H_{\beta} 
\end{equation}

\begin{displaymath}
\rm \Delta \log{\left( \frac{Z}{\Zsun} \right)} = 3.9738 \times 
                      \delta Mg_2 + 0.3026 \times \delta \langle Fe \rangle
\end{displaymath}
\begin{equation}
\rm ~~~~~~~~~ + 0.4687 \times \delta H_{\beta} 
\end{equation}

\begin{displaymath}
\rm \Delta \log(t) = - 2.5899 \times \delta Mg_2 - 0.0396 \times \delta \langle
                     Fe \rangle
\end{displaymath}
\begin{equation}
\rm ~~~~~~~~~ - 1.0424 \times \delta H_{\beta} 
\end{equation}
\noindent
The solutions of these equations are known with the following uncertainty: 
$\rm \Delta [Mg/Fe] \pm 0.0890$, $\rm \Delta log(Z/\Zsun) \pm 0.1524$,
and $\rm \Delta log(t) \pm 0.0538$. 
However, the analysis below will make use of the first original
formulation  because of  its better accuracy.

\begin{figure}
\psfig{file=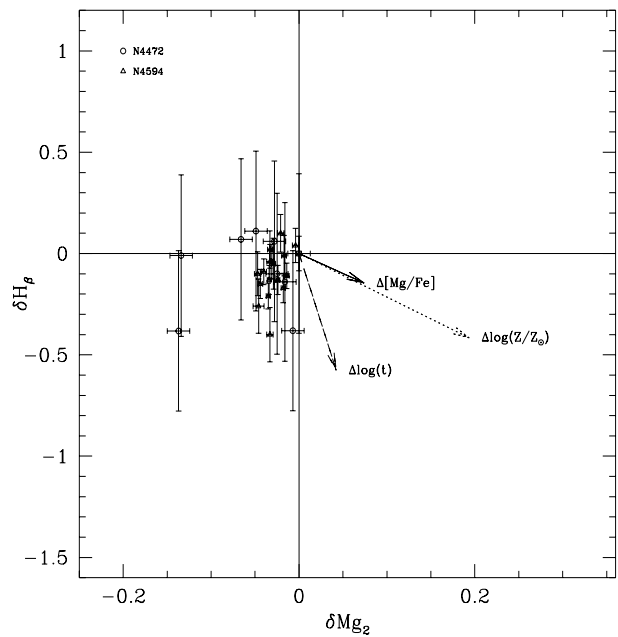,height=9.0truecm,width=8.5truecm}
\caption{The \dhb\ versus $\rm \delta Mg_{2}$ relation across individual galaxies. 
The three arrows
centered on (0,0) indicate the {\it age, metallicity, and enhancement vectors}
as indicated. Along the age vector, the age goes from 5 to 15 Gyr, along the
metallicity vector Z ranges from 0.004 to 0.05, and along the enhancement
vector  [Mg/Fe] goes from $-0.3$ to 0.4 dex. 
The data are from Vazdekis (1997, private communication) for NGC~4594,
and from Davies et al. (1993) for
NGC~4472.}
\label{dlog1}
\end{figure}

\section {Gradients across individual galaxies}
In this section we study the gradients in \Hbeta, \Mg2, and \MFe\ 
across a few individual galaxies. The analysis is made in two steps. 

We start showing in Fig.~\ref{dlog} the data for four galaxies taken from
different sources,  namely NGC~4472 from Davies et al. (1993), 
 NGC~4594 from Vazdekis (1997, private communication), and NGC~2434 and
NGC~3706 from Carollo \& Danziger (1994a,b). Galaxies have been chosen
on the base of the following criteria: elliptical objects and enough data 
per galaxy. This first step is meant to illustrate the method. 

The displayed data are the difference between the local value of
each index (at any radial distance) and its value at the center. Taking the
galaxy NGC~4594 as a prototype, we first derive an eye estimate of the mean
slope of the data, second we translate  the origin of the vectors to some
arbitrary external point of the galaxy. The result is that with respect to its
external regions, the nucleus is either more metal-rich and  older or more
metal-rich and slightly younger, and in any case more enhanced in [Mg/Fe]. It
goes without saying that if the nucleus is older than the periphery a lower
increase in the metallicity toward the center is required than in the opposite
alternative (younger nucleus). In such a case the increase in metallicity must
be large enough to compensate for the opposite trend of the age. Similar
considerations apply also to the other galaxies in Fig.~\ref{dlog}.

{\it How this result conforms to the  Bressan et al. (1996)  analysis of
Gonz\'ales (1993) galaxies in the \Hbeta\ and [MgFe], suggesting that in most
galaxies the nucleus is younger (star formation lasted longer) and more
metal-rich than the external regions, with a minority having the nucleus 
older and more metal-rich ?}

\begin{table*}
\begin{center}
\caption{Estimated gradients 
in age (Gyr), metallicity ($Z/\Zsun$) and $\alpha$-enhancement
across  a small group of galaxies. (1) Carollo \& Danziger (1994a); (2)
Davies et al. (1993); (3) Vazdekis (1997). See the text for more details. }
\begin{tabular*}{105mm}{c| c c c c c c c }
\hline
\hline
 & & & & & & &  \\
\multicolumn{1}{c|}{NGC} &
\multicolumn{1}{c}{Ref}&
\multicolumn{1}{c}{$\rm { {Mg_2}_N}$  } &
\multicolumn{1}{c}{$\rm { {\langle Fe \rangle}_N }$   } &
\multicolumn{1}{c}{$\rm { {H_{\beta}}_N  }$  } &
\multicolumn{1}{c}{$\rm { {Mg_2}_E}$    } &
\multicolumn{1}{c}{$\rm { {\langle Fe \rangle}_E }$   } &
\multicolumn{1}{c}{$\rm { {H_{\beta}}_E  }$  }\\
& & & & & & &  \\ 
\hline
& & & & & & & \\
2434  & (1)& 0.235  &  2.579  &  2.200   &  0.183 &   2.378 &   1.800\\
3706  & (1)& 0.292  &  2.937  &  1.800   &  0.232 &   2.584 &   1.800\\
3379  & (2)& 0.311  &  2.947  &  1.483   &  0.277 &   2.748 &   1.632\\
4374  & (2)& 0.318  &  2.900  &  1.263   &  0.289 &   2.683 &   1.494\\
4472  & (2)& 0.345  &  3.440  &  1.510   &  0.321 &   3.247 &   1.412\\
4594  & (3)& 0.332  &  3.102  &  1.183   &  0.212 &   1.974 &   0.943\\
 & & & & & & & \\
\hline
 & & & & & & & \\
\multicolumn{1}{l|}{NGC} &
\multicolumn{1}{c}{Ref } &
\multicolumn{1}{c}{\dmg  } &
\multicolumn{1}{c}{\dfem } &
\multicolumn{1}{c}{\dhb } &
\multicolumn{1}{c}{$\rm \Delta [{Mg\over Fe}]$  } &
\multicolumn{1}{c}{$\rm \Delta \log({Z\over Z_{\odot} }) $ } &
\multicolumn{1}{c}{$\rm \Delta \log(t) $}\\
& & & & & & &  \\
\hline 
& & & & & & & \\
2434  &(1) &-0.052 &  -0.201 & -0.400 & -0.231 & -0.409 &  0.500\\
3706  &(1) &-0.060 &  -0.353 &  0.000 & -0.154 & -0.311 &  0.124\\
3379  &(2) &-0.034 &  -0.199 &  0.149 & -0.070 & -0.112 & -0.076\\
4374  &(2) &-0.029 &  -0.217 &  0.231 & -0.025 & -0.066 & -0.167\\
4472  &(2) &-0.024 &  -0.193 & -0.098 & -0.048 & -0.181 &  0.146\\
4594  &(3) &-0.120 &  -1.146 & -0.240 & -0.126 & -0.851 &  0.488\\
 & & & & & & & \\
\hline
\hline
\end{tabular*}
\end{center}
\normalsize
\label{tab5}
\end{table*}

\begin{figure}
\psfig{file=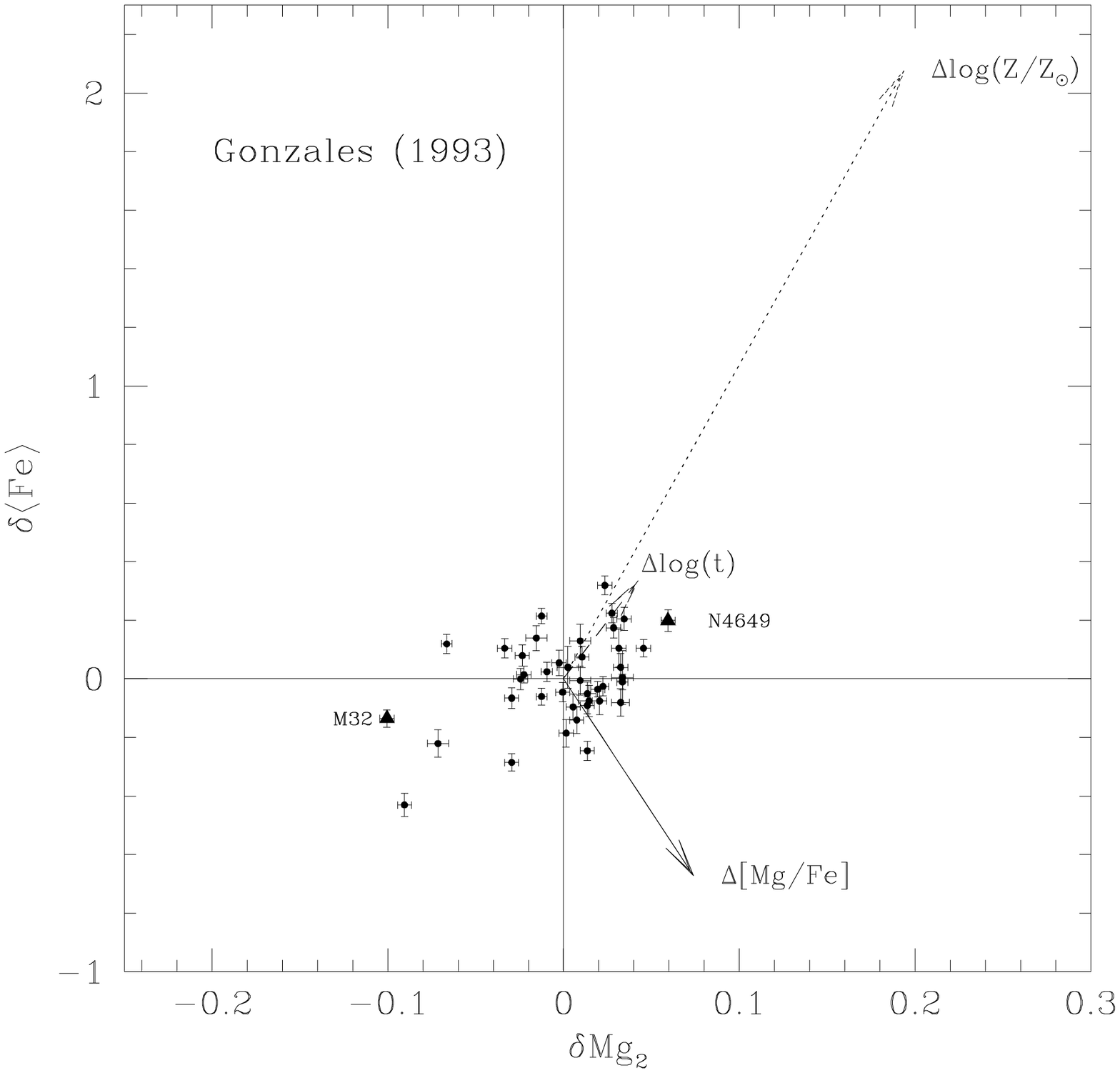,height=9.0truecm,width=8.5truecm}
\caption{The \dfem\ versus $\rm \delta Mg_{2}$
relation. The three arrows centered on (0,0) are the {\it age, metallicity,
and enhancement vectors} as indicated. They have been calculated as in Fig.1.
The displayed data  are for each galaxy the difference between its central
value and the mean value of the sample. The position of M32 and  NGC~4649 is indicated. }
\label{dgal_nuc}
\end{figure}

\begin{figure}
\psfig{file=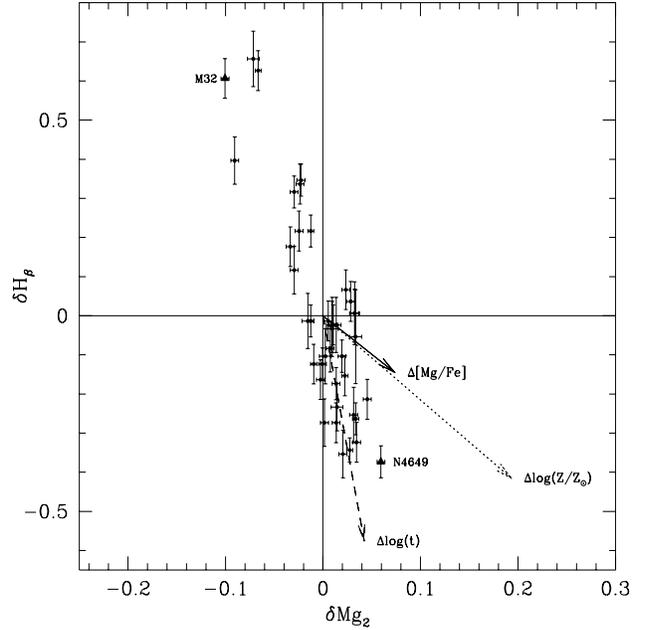,height=9.0truecm,width=8.5truecm}
\caption{The \dhb\ versus $\rm \delta Mg_{2}$ relation. The
three arrows centered on (0,0) are the {\it age, metallicity  and enhancement
vectors} as indicated. They have been calculated as in Fig.2. The data are
from Gonz\'ales (1993) and for each galaxy the difference between its central
values and the mean value of the sample are displayed. The position of M32
and NGC~4649 is indicated.}
\label{dgal_nucHb}
\end{figure}

To answer the above question we make use of the \Hbeta\ index, known
to be more sensitive to the age, and look at the parameter space $\rm
H_{\beta}$, \Mg2\ and \MFe. 

In Fig.~\ref{dlog1} the same type of data as in Fig.~\ref{dlog} is
shown,  but in the \dhb\ versus \dmg\ plane, together with the
theoretical age, metallicity, and [Mg/Fe] vectors. Now the number of
galaxies  that are plotted is very small, only two objects indeed,
namely NGC~4472 from Davies et al. ( 1993) and NGC~4594 from Vazdekis (1997,
private communication). Two more galaxies from Davies et al. (1993) are not
shown on Fig.~\ref{dlog1} for the sake of clarity.

We learn from this diagram that the two galaxies in question have their 
nuclear region containing stars that are more enhanced in \alfa, more 
metal-rich, and younger than the external regions.

To strengthen the above preliminary conclusion  we have repeated the analysis
by means of the $\Delta$-method. The galaxies under examinations are those 
already displayed in Figs.~\ref{dlog} and \ref{dlog1} plus NGC~3379 and NGC~4374 
taken from Davies et al. (1993). The results are summarized in Table~2. The top 
part of the table shows the observational data, i.e. the mean values of \Mg2, 
$\rm \langle Fe \rangle$, and \Hbeta\ of the central region (indicated by a capitol N) 
and the external region (labelled by a capitol E). The central region is defined by 
visually inspecting the  \Mg2\  gradient in the case of the Carollo \& Danziger (1994a,b) 
objects (5 arcsec from the galactic center) or taken according to the definition given by the
authors for the remaining galaxies.

The bottom part of Table~2 lists the difference $\rm \delta Mg_2$, \dfem, and
\dhb\ between the external and central values of the three indices, and the
solution found for  $\rm \Delta [Mg/Fe]$, $\rm \Delta \log(Z/Z_{\odot})$, and 
\Dlogt\ between the periphery and center.

It turns out that of the six  galaxies, four have the nuclear region
containing  stars more enhanced in \alfa, more metal-rich, and younger
than the external regions,  whereas two of them have the nucleus more 
metal-rich, more $\alpha$-enhanced but (marginally) older than the 
external regions.

\section{Going from galaxy to galaxy}
 Given that the central values of \Hbeta, \Mg2, and $\rm \langle Fe \rangle$ are 
known to vary from galaxy to galaxy, which are the implications are as far as the 
differences in age, metallicity, and enhancement in \alfa\  are concerned? 

To answer this question we make use of the Gonz\'ales (1993) sample 
of elliptical galaxies (41 objects in total) containing the so-called
{\it nuclear data}, i.e. within $2\times 4.1$ arcsec. The data are presented 
in Table~3. Column (1) identifies the galaxy: those objects whose name is followed 
by a V belong to the Virgo cluster. Columns (2) to (5) give \Mg2, $\rm Fe_{5270}$, $\rm
Fe_{5335}$, and \Hbeta. Columns (6), (7) and (8) show the differences \dhb, 
$\rm \delta Mg_{2}$, and $\rm \delta \langle Fe \rangle$ of the galaxy indices 
with respect to the mean values (see below), respectively. Columns (9), (10), and 
(11) give the difference with respect to the mean of the enrichment factor $\rm \Delta
[Mg/Fe]$, metallicity $\rm \Delta \log(Z/Z_{\odot})$, and age $\rm \Delta
\log(t)$ found for each galaxy. Columns (12) and (13) lists the total absolute
blue and visual magnitudes $M_B$ and $M_V$, respectively. Finally, column 
(14) lists the rejuvenation parameter $\Sigma$ defined by Schweizer \& Seitzer 
(1992). For each galaxy, the top raw are the data (either observational or 
theoretical), whereas the bottom raw are the corresponding uncertainties.

To perform the analysis  we have calculated the mean values of \Mg2, $\rm
\langle Fe \rangle$ and \Hbeta\ for the whole sample: $\rm \overline{H_{\beta}}=1.75$, 
$\rm \overline{Mg_{2}}=0.32$, and  $\rm \overline{\langle Fe \rangle}=2.96$.

They are used as the origin of a new system of coordinates in which the differences 
$\rm \delta Mg_{2}$, $\rm \delta \langle Fe \rangle$ and \dhb\ passing from galaxy to
galaxy are expressed.

\subsection{ The $\rm \delta Mg_{2}$-\dfem\ and   
 $\rm \delta Mg_{2}$-\dhb\ planes}

The observational data for the differences $\rm \delta  Mg_{2}$ and $\rm
\delta \langle Fe \rangle$ are shown in Fig.~\ref{dgal_nuc}, in which we have
drawn the metallicity, enhancement, and age  vectors, whose length is the same
as in the previous analysis, and finally indicated the positions of M32 and
NGC~4649, the prototype galaxies in the discussion below. Once again, age 
and metallicity cannot be safely separated, whereas this is feasible for [Mg/Fe].

In order to cope with the age-metallicity degeneracy encountered in the  $\rm
\delta Mg_{2}$ - \dfem\ plane, we look at the $\rm \delta Mg_{2}$ - \dhb\ 
plane owing to the better age sensitivity of \Hbeta. This plane is shown in 
Fig.~\ref{dgal_nucHb} where the module of the three vectors are:
$\rm \Delta \log(Z) \sim 1.1$, $\Delta \rm [Mg/Fe] \sim 0.7$ and  $\rm \Delta
\log(t) \sim 0.47$. In this figure the effect of [Mg/Fe] and metallicity
cannot be isolated because the two vectors are coincident (degeneracy),
whereas this is feasible for the age.

With the aid of the $\Delta$-method  and the differences \dhb, \dmg, and $\rm \delta
\langle Fe \rangle$, we derive for each galaxy the variations in age $\rm
\Delta \log(t)$, metallicity $\rm \Delta \log(Z/Z_{\odot})$,  and enhancement
$\rm \Delta [Mg/Fe]$ with respect to the mean values. The results together
with the expected uncertainties  are listed in columns (9), (10) and (11) of Table~3.

Looking at M32 and NGC~4649 as an example, we get the following results: for
M32 $\rm \Delta \log(t)=-0.392$, $\rm \Delta \log(Z/Z_{\odot})=-0.131$, and
$\rm \Delta [Mg/Fe]=-0.403$; for NGC~4649 $\rm \Delta \log(t)=0.245$, $\rm
\Delta \log(Z/Z_{\odot})=0.106$, and $\rm \Delta [Mg/Fe]=0.180$ with respect 
to the mean value. The stellar content of M32 is less metal rich, less enhanced 
in \alfa, and much younger (or more safely has a much younger component contributing 
to $\rm H_{\beta}$) than NGC~4649.

\begin{figure}
\psfig{file=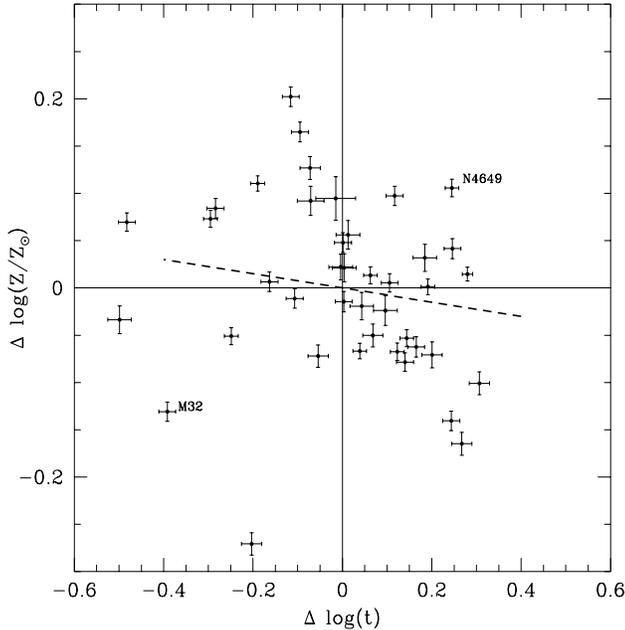,height=9.0truecm,width=8.5truecm}
\caption{The $\rm \Delta \log(Z/Z_{\odot})$ versus \Dlogt\
relation. The {\em thick dashed line} is the linear regression of the results.
The position of M32 and NGC~4649 is indicated.}
\label{log_t_z}
\end{figure}

\begin{figure}
\psfig{file=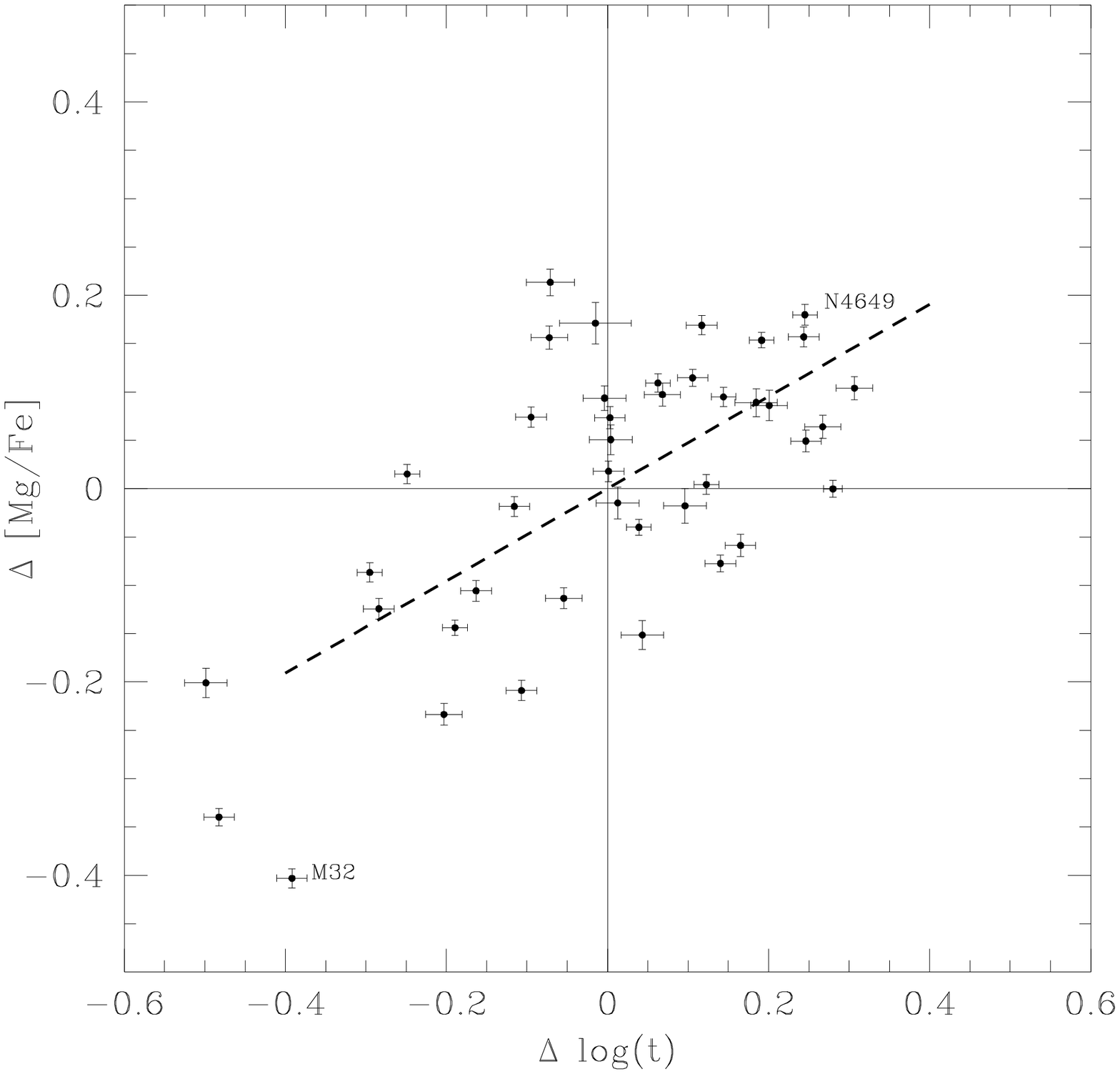,height=9.0truecm,width=8.5truecm}
\caption{The $\rm \Delta [Mg/Fe])$ versus \Dlogt\ relation. The
{\em thick dashed line} is the linear regression of the results. The position
of M32 and NGC~4649 is indicated.}
\label{log_t_mgfe}
\end{figure}

\begin{figure}
\psfig{file=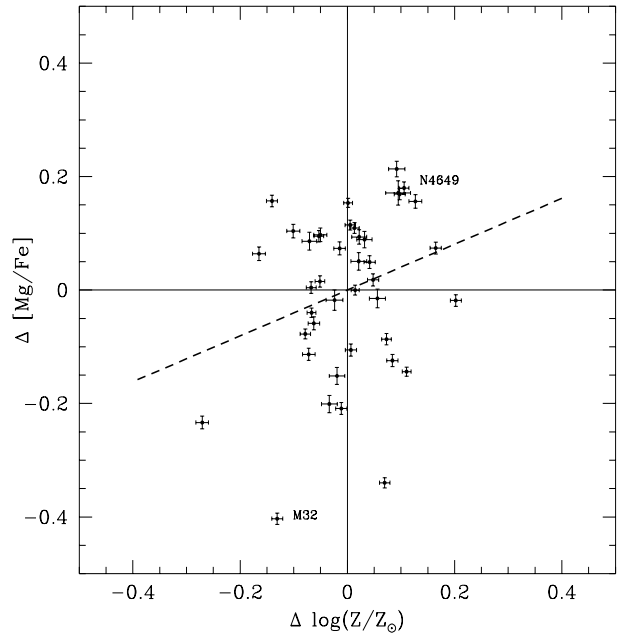,height=9.0truecm,width=8.5truecm}
\caption{The $\rm \Delta [Mg/Fe]$ versus $\rm \Delta \log(Z/Z_{\odot})$
relation. The {\em thick dashed line} is the linear regression of the results.
The position of M32 and NGC~4649 is indicated.}
\label{log_z_mgfe}
\end{figure}

\section{ The \Dlogt, $\rm \Delta \log(Z/Z_{\odot})$, 
             $\rm \Delta [Mg/Fe]$ space}

Aim of this section is to investigate whether systematic correlations exist
among the three physical quantities  \Dlogt, $\rm \Delta
\log(Z/Z_{\odot})$, and $\rm \Delta [Mg/Fe]$. The occurrence of such relations
would bear very much on the past history of star formation and chemical
enrichment, and the mechanism of galaxy formation as well.

Before starting this part of the analysis it is worth clarifying the real
meaning of the {\it age } parameter  \Dlogt. This age (being
mainly revealed by \Hbeta) actually refers to the younger components of
the stellar mix in the galaxy (nucleus). In the ideal case (however possible)
of continuous star formation from the early epochs it would measure the
overall duration of the star forming activity. In the case that recurrent
episodes have taken place, \Hbeta\  would measure the age of the last episode. 
Given these premises, we examine the following relationships:

\littleskip
{\it Age-Z}. The relation between  $\rm \Delta \log(Z/Z_{\odot})$ and $\rm
\Delta \log(t)$ is shown in Fig.~\ref{log_t_z}, where only a large scatter is
seen (no systematic trend). The degree of metal enrichment seems to be totally
unrelated to the age, in the sense that at any given metallicity all ages are
possible. Does it means that sporadic episodes of star formation may change
the age parameter without significantly affecting the metallicity? It is worth
recalling that \Hbeta\ is very sensitive to recent star formation, and
that a burst of stellar activity implying even a very small fraction of the
galaxy mass would immediately increase \Hbeta\ without affecting the
metallicity (cf. Bressan et al. 1996). Furthermore, the recovery time of $\rm
H_{\beta}$ after a burst is of order of about 1 Gyr (with some dependence on the 
amount of mass involved in the star forming episode), after which  no trace of
the star forming period would be easily detectable with the \Hbeta\
diagnostic.
\littleskip

{\it Age-[Mg/Fe]}. The relationship between $\rm \Delta [Mg/Fe]$ and $\rm
\Delta \log(t)$ is shown in Fig.~\ref{log_t_mgfe}. Now a good relation is
found: older galaxies seem to be more enhanced in \alfa.
\littleskip

{\it Z-[Mg/Fe]}. The relationship between $\rm \Delta [Mg/Fe]$ and $\rm \Delta
\log(Z/Z_{\odot})$ is shown in Fig.~\ref{log_z_mgfe}. It appears that more
metal-rich  galaxies are likely also more enhanced in \alfa.
\littleskip

\begin{figure}
\psfig{file=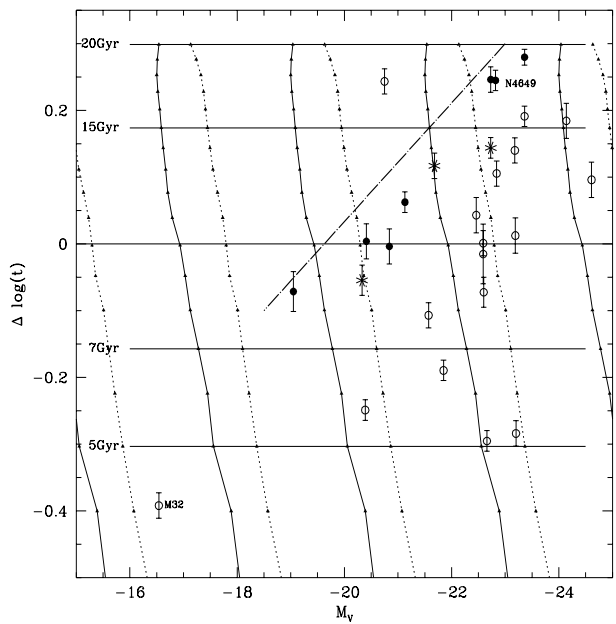,height=9.0truecm,width=8.5truecm}
\caption{The \Dlogt\ versus $\rm M_V$ relation. Filled circles are the galaxies 
defining the age limit (see the text for details), the heavy stars are the three 
galaxies of this group belonging to the Virgo cluster, and the open circles are 
all remaining galaxies. The {\em eye drawn long-dashed-dotted line} is only meant 
to visualize the age limit. Finally, the position of M32, NGC~4649 is indicated.
The fading lines of SSPs are superposed: the thin dotted and solid lines are for 
Z=0.004 and Z=0.05, respectively; each dot along the lines shows the age in step of 
1 Gyr starting from 20 Gyr (top) down to 4 Gyr (bottom); the thin horizontal lines
locate the loci of constant age (20, 15, 10, 7, and 5 Gyr starting from the
top). Finally, the fading lines of SSP (in pairs because of the different
metallicity) are shown for different values of the total mass, namely
$10^{12}$, $10^{11}$, $10^{10}$, $10^9$, and $10^8\rm \times M_{\odot}$ from
right to left.}
\label{age_mv}
\end{figure}

\begin{figure}
\psfig{file=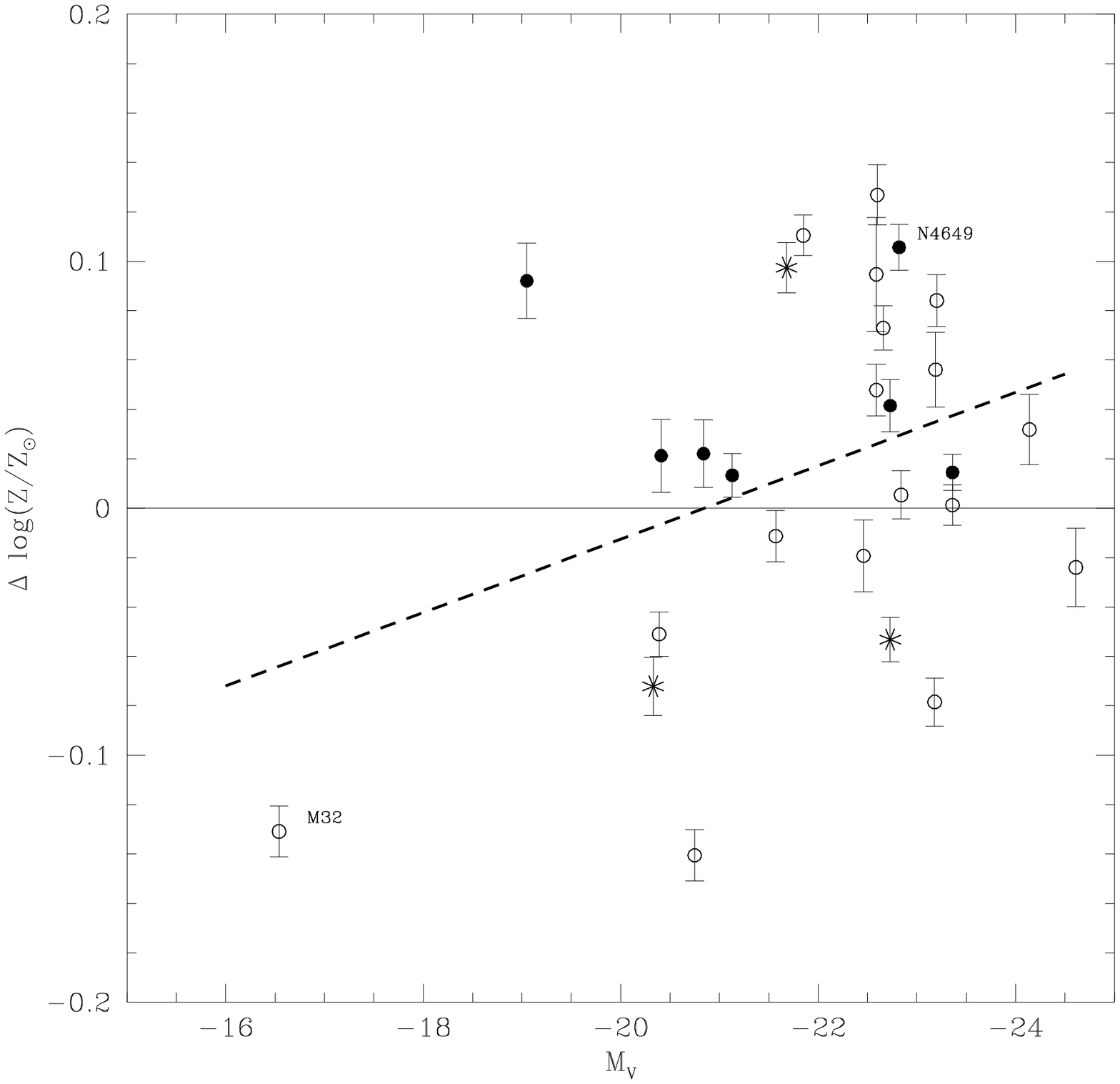,height=9.0truecm,width=8.5truecm}
\caption{The $\rm \Delta \log(Z/Z_{\odot})$ versus $\rm M_V$ relation. The
{\em thick dashed line} is the linear regression of the results. 
Filled circles are the galaxies defining the age limit (see the text for
details), the heavy stars are the three galaxies of this group belonging
to the Virgo cluster, and the open circles are all remaining galaxies.  
 Finally, the position of M32,
NGC~4649 is indicated.}
\label{zeta_mv}
\end{figure}

\begin{figure}
\psfig{file=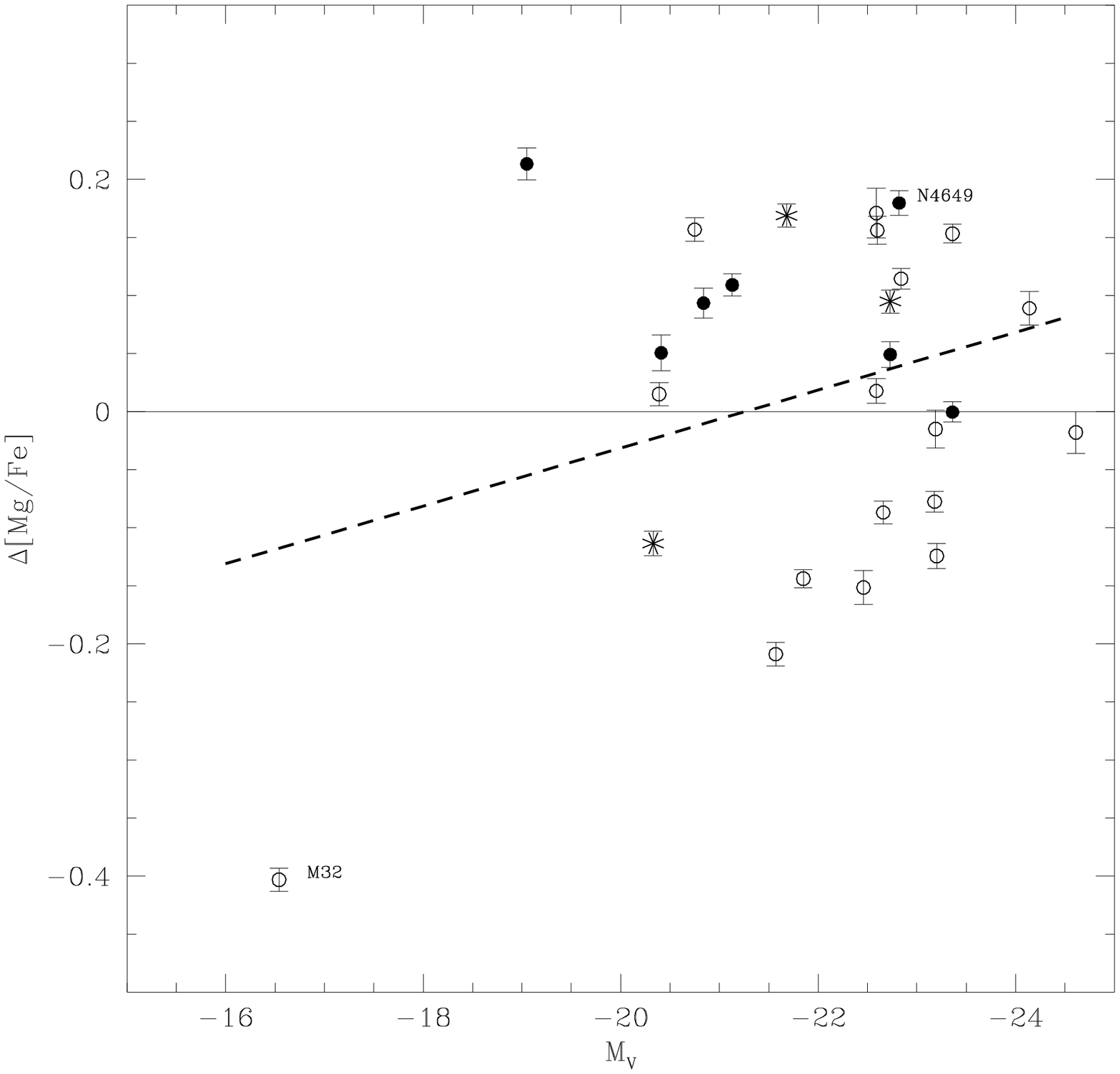,height=9.0truecm,width=8.5truecm}
\caption{The $\rm \Delta [Mg/Fe]$ versus $\rm M_V$ relation. The {\em thick
dashed line} is the linear regression of the results. 
Filled circles are the galaxies defining the age limit (see the text for
details), the heavy stars are the three galaxies of this group belonging
to the Virgo cluster, and the open circles are all remaining galaxies.  
 Finally, the position of M32, NGC~4649 is indicated.}
\label{mgfe_mv}
\end{figure}

In order to cast light on the physical implications of the above relations, we
separately correlate \Dlogt, $\rm \Delta \log(Z/Z_{\odot}$), and $\rm \Delta 
[Mg/Fe]$ to the total luminosity of the galaxy as measured by the absolute 
magnitude $M_V$. The discussion below will not change using $\rm M_B$. The 
three relations are shown in Fig.~\ref{age_mv} (age), \ref{zeta_mv} (metallicity), 
and \ref{mgfe_mv} ([Mg/Fe]). 
\littleskip

{\it Age-magnitude plane: \Dlogt-$M_V$}. The distribution of our galaxies
in the \Dlogt-$M_V$ plane is shown in Fig.~\ref{age_mv} together with the 
fading lines of SSPs of different composition, i.e. Z=0.004 (dotted lines) 
and Z=0.05 (solid lines). Each dot along the lines shows the age in step of 
1 Gyr  starting from 20 Gyr (top) down to 4 Gyr (bottom); the thin horizontal 
lines locate the loci of constant age (20, 15, 10, 7, and
5 Gyr starting from the top). Finally, the fading lines of SSP (in pairs
because of the different metallicity) are shown for different values of the
total mass, namely $10^{12}$,  $10^{11}$,  $10^{10}$, $10^9$, and $10^8 \times
M_{\odot}$  from right to left. Mounting  the two sets of data, namely the age
differences \Dlogt\ of galaxies and the absolute ages  SSPs, is made assigning 
the last star forming episode in M32 the provisional age of $4\div5$. See the 
more detailed discussion below.

With these premises, the following features can be noticed in Fig.~\ref{age_mv} 
despite the large scatter and the small number of  objects: 
\begin{itemize}

\item{In our sample all galaxies but  M32 are brighter than 
$\rm M_V=-20$, which is the range of the classical color-magnitude relation 
of Bower et al. (1992). Their corresponding masses fall  in the range 
$10^{10} \div 10^{12}\times M_{\odot}$. }
\littleskip

\item{Leaving M32 aside, there seems to be a sort of upper limit to \Dlogt\ (and absolute
age in turn) set by the  galaxies NGC~2778, NGC~3818, NGC~3379, NGC~4552, 
NGC~4649, NGC~4261, and NGC~7619, to which NGC~4478, NGC~3608 and NGC~4374 
can perhaps be added. There is one exception though, represented by 
NGC~4278, which appears to be either too old for its luminosity or too 
faint for its age. }
\littleskip

\item {Three galaxies of this group, namely NGC~4374, NGC~4478, and NGC~4552, 
belong to the Virgo cluster (so that their distance and magnitude are less of 
a problem in this context) and have $\Sigma \leq 2$, which according to Sweitzer 
\& Seitzer (1992) means mild rejuvenation and/or interaction. The remaining galaxies 
of this sub-group have either $\Sigma \leq 1.5$ or missing. Finally, the last Virgo 
galaxy in our sample, namely NGC~4697, deviates from the relation defined by its 
companions in spite of its similar $\rm M_V$ and no signs of interaction (Sweitzer and
Seitzer 1992).}
\end{itemize}

\begin{table*}
\begin{center}
\caption{Nuclear  values of \Hbeta, \Mg2\ and \MFe\ and solutions
$\rm \Delta log(t)$, $\rm \Delta log(Z/\Zsun)$, and $\rm \Delta [Mg/Fe]$ for 
the galaxies of Gonz\'ales (1993). The mean observational values are
$\rm \overline{H_{\beta}}=1.75 \pm 0.009$, 
$\rm \overline{Mg_{2}}=0.32 \pm 0.001$, and 
$\rm \overline{\langle Fe \rangle}=2.96 \pm 0.006$. }
\scriptsize
\begin{tabular*}{170mm}{l| c c c c c c r r r r r r c }
\hline
\hline
 & & & & & & & & & & & & & \\
\multicolumn{1}{l|}{NGC} &
\multicolumn{1}{c}{\Mg2} &
\multicolumn{1}{c}{$\rm Fe_{52}$} &
\multicolumn{1}{c}{$\rm Fe_{53}$} &
\multicolumn{1}{c}{$\rm H_{\beta}$} &
\multicolumn{1}{c}{\dmg } &
\multicolumn{1}{c}{\dhb } &
\multicolumn{1}{c}{$\rm \delta \langle Fe \rangle  $} &
\multicolumn{1}{c}{$\rm \Delta [{Mg \over Fe}]$} &
\multicolumn{1}{c}{$\rm \Delta \log({Z\over Z_{\odot} }) $} &
\multicolumn{1}{c}{\Dlogt } &
\multicolumn{1}{c}{$\rm M_B $ } &
\multicolumn{1}{c}{$\rm M_V $ } &
\multicolumn{1}{c}{$\Sigma$ }\\
 & & & & & & & & & & & & &\\
\hline
 & & & & & & & & & & & & &\\
221  & 0.216 & 2.99 & 2.66 & 2.36 & -0.1005 & -0.1357 &  0.6066 & -0.4030 & -0.1309 & -0.3919 & -15.70 & -16.54 &  -   \\
     & 0.004 & 0.04 & 0.04 & 0.05 &  0.004  &  0.029  &  0.051  &  0.0100 &  0.0102 &  0.0190 &        &        &      \\
224  & 0.340 & 3.39 & 3.17 & 1.82 &  0.0235 &  0.3193 &  0.0666 & -0.0184 &  0.2022 & -0.1156 &    -   &    -   &  -   \\
     & 0.004 & 0.04 & 0.05 & 0.05 &  0.004  &  0.033  &  0.051  &  0.0103 &  0.0103 &  0.0190 &        &        &      \\
315  & 0.322 & 2.89 & 2.84 & 1.74 &  0.0055 & -0.0957 & -0.0134 &  0.0734 & -0.0145 &  0.0030 &    -   &    -   &  -   \\
     & 0.004 & 0.05 & 0.07 & 0.05 &  0.004  &  0.043  &  0.051  &  0.0115 &  0.0108 &  0.0190 &        &        &      \\
507  & 0.319 & 3.01 & 2.99 & 1.65 &  0.0025 &  0.0393 & -0.1034 & -0.0177 & -0.0240 &  0.0963 &    -   & -24.61 &  -   \\
     & 0.006 & 0.09 & 0.11 & 0.07 &  0.006  &  0.071  &  0.071  &  0.0179 &  0.0158 &  0.0265 &        &        &      \\
547  & 0.324 & 2.98 & 2.66 & 1.67 &  0.0075 & -0.1407 & -0.0834 &  0.0973 & -0.0503 &  0.0682 &    -   &    -   &  -   \\
     & 0.004 & 0.06 & 0.07 & 0.06 &  0.004  &  0.046  &  0.061  &  0.0120 &  0.0122 &  0.0226 &        &        &      \\
584  & 0.294 & 3.11 & 2.84 & 2.10 & -0.0225 &  0.0143 &  0.3466 & -0.0867 &  0.0730 & -0.2951 & -21.72 & -22.66 & 2.78 \\
     & 0.004 & 0.04 & 0.03 & 0.04 &  0.004  &  0.029  &  0.041  &  0.0099 &  0.0190 &  0.0155 &        &        &      \\
636  & 0.283 & 3.24 & 2.89 & 1.93 & -0.0335 &  0.1043 &  0.1766 & -0.2088 & -0.0113 & -0.1068 & -20.65 & -21.57 & 1.48 \\
     & 0.004 & 0.04 & 0.05 & 0.05 &  0.004  &  0.033  &  0.051  &  0.0103 &  0.0103 &  0.0190 &        &        &      \\
720  & 0.350 & 3.03 & 2.90 & 1.70 &  0.0335 &  0.0043 & -0.0534 &  0.1712 &  0.0947 & -0.0150 & -21.60 & -22.59 &  -   \\
     & 0.006 & 0.12 & 0.14 & 0.12 &  0.006  &  0.092  &  0.120  &  0.0212 &  0.0231 &  0.0447 &        &        &      \\
821  & 0.327 & 3.20 & 2.87 & 1.73 &  0.0105 &  0.0743 & -0.0234 &  0.0180 &  0.0479 &  0.0012 & -21.61 & -22.59 &  -   \\
     & 0.004 & 0.05 & 0.05 & 0.05 &  0.004  &  0.036  &  0.051  &  0.0107 &  0.0105 &  0.0190 &        &        &      \\
1453 & 0.314 & 3.05 & 2.98 & 1.59 & -0.0025 &  0.0543 & -0.1634 & -0.0588 & -0.0624 &  0.1651 &    -   &    -   & 1.48 \\
     & 0.004 & 0.05 & 0.07 & 0.05 &  0.004  &  0.043  &  0.051  &  0.0115 &  0.0108 &  0.0190 &        &        &      \\
1600 & 0.348 & 2.95 & 3.18 & 1.50 &  0.0315 &  0.1043 & -0.2534 &  0.0891 &  0.0319 &  0.1845 & -23.17 & -24.14 &  -   \\
     & 0.004 & 0.08 & 0.10 & 0.07 &  0.004  &  0.064  &  0.071  &  0.0145 &  0.0143 &  0.0263 &        &        &      \\
1700 & 0.293 & 3.20 & 2.88 & 2.09 & -0.0235 &  0.0793 &  0.3366 & -0.1242 &  0.0841 & -0.2839 & -22.28 & -23.20 & 3.70 \\
     & 0.004 & 0.05 & 0.05 & 0.05 &  0.004  &  0.036  &  0.051  &  0.0107 &  0.0105 &  0.0190 &        &        &      \\
2300 & 0.349 & 3.15 & 2.85 & 1.76 &  0.0325 &  0.0393 &  0.0066 &  0.1563 &  0.1269 & -0.0723 & -21.56 & -22.60 & 2.85 \\
     & 0.004 & 0.06 & 0.07 & 0.06 &  0.004  &  0.046  &  0.061  &  0.0120 &  0.0122 &  0.0226 &        &        &      \\
2778 & 0.349 & 3.05 & 2.71 & 1.76 &  0.0325 & -0.0807 &  0.0066 &  0.2135 &  0.0921 & -0.0712 & -18.14 & -19.05 &  -   \\
     & 0.005 & 0.06 & 0.07 & 0.08 &  0.005  &  0.046  &  0.080  &  0.0137 &  0.0153 &  0.0300 &        &        &      \\
3377 & 0.287 & 2.82 & 2.53 & 2.07 & -0.0295 & -0.2857 &  0.3166 &  0.0152 & -0.0510 & -0.2488 & -19.49 & -20.39 & 1.48 \\
     & 0.004 & 0.04 & 0.04 & 0.04 &  0.004  &  0.029  &  0.041  &  0.0099 &  0.0090 &  0.0155 &        &        &      \\
3379 & 0.336 & 3.07 & 2.78 & 1.65 &  0.0195 & -0.0357 & -0.1034 &  0.1092 &  0.0133 &  0.0627 & -20.17 & -21.13 & 0.00 \\
     & 0.004 & 0.03 & 0.04 & 0.04 &  0.004  &  0.026  &  0.041  &  0.0096 &  0.0089 &  0.0155 &        &        &      \\
3608 & 0.330 & 3.00 & 2.82 & 1.73 &  0.0135 & -0.0507 & -0.0234 &  0.0937 &  0.0221 & -0.0037 & -19.87 & -20.84 & 0.00 \\
     & 0.005 & 0.05 & 0.06 & 0.07 &  0.005  &  0.039  &  0.071  &  0.0128 &  0.0138 &  0.0264 &        &        &      \\
3818 & 0.326 & 3.05 & 2.86 & 1.73 &  0.0095 & -0.0057 & -0.0234 &  0.0508 &  0.0213 &  0.0040 & -19.49 & -20.41 & 1.30 \\
     & 0.006 & 0.07 & 0.07 & 0.07 &  0.006  &  0.050  &  0.071  &  0.0154 &  0.0148 &  0.0265 &        &        &      \\
4261 & 0.351 & 3.29 & 3.04 & 1.43 &  0.0345 &  0.2043 & -0.3234 &  0.0493 &  0.0416 &  0.2464 & -21.74 & -22.73 & 1.00 \\
     & 0.004 & 0.05 & 0.06 & 0.05 &  0.004  &  0.039  &  0.051  &  0.0111 &  0.0106 &  0.0190 &        &        &      \\
4278 & 0.330 & 2.76 & 2.67 & 1.48 &  0.0135 & -0.2457 & -0.2734 &  0.1570 & -0.1405 &  0.2436 & -19.79 & -20.75 & 1.48 \\
     & 0.004 & 0.04 & 0.05 & 0.05 &  0.004  &  0.033  &  0.051  &  0.0103 &  0.0103 &  0.0190 &        &        &      \\
4374~V & 0.330 & 3.01 & 2.73 & 1.58 &  0.0135 & -0.0907 & -0.1734 &  0.0950 & -0.0532 &  0.1440 & -21.85 & -22.73 & 2.30 \\
     & 0.004 & 0.04 & 0.04 & 0.04 &  0.004  &  0.029  &  0.041  &  0.0099 &  0.0090 &  0.0155 &        &        &      \\
4472~V & 0.326 & 3.03 & 3.15 & 1.72 &  0.0095 &  0.1293 & -0.0334 & -0.0148 &  0.0561 &  0.0126 & -22.21 & -23.19 &  -   \\
     & 0.006 & 0.07 & 0.09 & 0.07 &  0.006  &  0.057  &  0.071  &  0.0163 &  0.0151 &  0.0265 &        &        &      \\
4478~V & 0.287 & 2.99 & 2.80 & 1.87 & -0.0295 & -0.0657 &  0.1166 & -0.1134 & -0.0721 & -0.0544 & -19.44 & -20.33 &  -   \\
     & 0.004 & 0.05 & 0.05 & 0.06 &  0.004  &  0.036  &  0.061  &  0.0108 &  0.0117 &  0.0226 &        &        &      \\
4489 & 0.245 & 2.99 & 2.49 & 2.41 & -0.0715 & -0.2207 &  0.6566 & -0.2009 & -0.0336 & -0.4987 &    -   &    -   &  -   \\
     & 0.006 & 0.06 & 0.07 & 0.07 &  0.006  &  0.046  &  0.071  &  0.0151 &  0.0147 &  0.0265 &        &        &      \\
4552~V & 0.362 & 3.07 & 3.06 & 1.54 &  0.0445 &  0.1043 & -0.2134 &  0.1690 &  0.0975 &  0.1170 &    -   & -21.68 & 2.00 \\
     & 0.004 & 0.04 & 0.04 & 0.05 &  0.004  &  0.029  &  0.051  &  0.0100 &  0.0102 &  0.0190 &        &        &      \\
4649 & 0.376 & 3.10 & 3.22 & 1.38 &  0.0595 &  0.1993 & -0.3734 &  0.1799 &  0.1057 &  0.2451 & -21.81 & -22.82 &  -   \\
     & 0.004 & 0.04 & 0.06 & 0.04 &  0.004  &  0.037  &  0.041  &  0.0107 &  0.0093 &  0.0155 &        &        &      \\
4697~V & 0.301 & 3.25 & 2.95 & 1.74 & -0.0155 &  0.1393 & -0.0134 & -0.1514 & -0.0193 &  0.0432 & -21.51 & -22.46 & 0.00 \\
     & 0.006 & 0.06 & 0.06 & 0.07 &  0.006  &  0.043  &  0.071  &  0.0148 &  0.0145 &  0.0265 &        &        &      \\
5638 & 0.316 & 3.12 & 2.71 & 1.63 & -0.0005 & -0.0457 & -0.1234 &  0.0043 & -0.0675 &  0.1227 &    -   &    -   &  -   \\
     & 0.004 & 0.04 & 0.05 & 0.04 &  0.004  &  0.033  &  0.041  &  0.0102 &  0.0091 &  0.0155 &        &        &      \\
5812 & 0.345 & 3.15 & 3.12 & 1.79 &  0.0285 &  0.1743 &  0.0366 &  0.0740 &  0.1648 & -0.0949 &    -   &    -   &  -   \\
     & 0.004 & 0.05 & 0.05 & 0.05 &  0.004  &  0.036  &  0.051  &  0.0107 &  0.0105 &  0.0190 &        &        &      \\
5813 & 0.318 & 2.95 & 2.60 & 1.48 &  0.0015 & -0.1857 & -0.2734 &  0.0640 & -0.1648 &  0.2673 &    -   &    -   &  -   \\
     & 0.004 & 0.06 & 0.07 & 0.06 &  0.004  &  0.046  &  0.061  &  0.0120 &  0.0122 &  0.0226 &        &        &      \\
5831 & 0.304 & 3.32 & 3.03 & 1.97 & -0.0125 &  0.2143 &  0.2166 & -0.1438 &  0.1105 & -0.1894 & -20.22 & -21.85 & 3.60 \\
     & 0.003 & 0.03 & 0.04 & 0.04 &  0.003  &  0.026  &  0.041  &  0.0080 &  0.0082 &  0.0153 &        &        &      \\
5846 & 0.339 & 3.03 & 2.84 & 1.60 &  0.0225 & -0.0257 & -0.1534 &  0.1147 &  0.0054 &  0.1056 & -21.85 & -22.84 & 0.30 \\
     & 0.003 & 0.04 & 0.05 & 0.05 &  0.003  &  0.033  &  0.051  &  0.0088 &  0.0097 &  0.0189 &        &        &      \\
6127 & 0.331 & 2.95 & 2.82 & 1.52 &  0.0145 & -0.0757 & -0.2334 &  0.0861 & -0.0708 &  0.2008 &    -   &    -   &  -   \\
     & 0.006 & 0.07 & 0.08 & 0.06 &  0.006  &  0.053  &  0.061  &  0.0158 &  0.0138 &  0.0229 &        &        &      \\
6702 & 0.250 & 3.19 & 2.97 & 2.38 & -0.0665 &  0.1193 &  0.6266 & -0.3397 &  0.0695 & -0.4824 &    -   &    -   &  -   \\
     & 0.003 & 0.04 & 0.05 & 0.05 &  0.003  &  0.033  &  0.051  &  0.0088 &  0.0097 &  0.0189 &        &        &      \\
6703 & 0.292 & 3.10 & 2.82 & 1.97 & -0.0245 & -0.0007 &  0.2166 & -0.1057 &  0.0065 & -0.1633 &    -   &    -   &  -   \\
     & 0.004 & 0.05 & 0.05 & 0.05 &  0.004  &  0.036  &  0.051  &  0.0107 &  0.0105 &  0.0190 &        &        &      \\
7052 & 0.337 & 2.95 & 2.82 & 1.40 &  0.0205 & -0.0757 & -0.3534 &  0.1041 & -0.1009 &  0.3066 &    -   &    -   &  -   \\
     & 0.004 & 0.06 & 0.07 & 0.06 &  0.004  &  0.046  &  0.061  &  0.0120 &  0.0122 &  0.0226 &        &        &      \\
7454 & 0.226 & 2.68 & 2.38 & 2.15 & -0.0905 & -0.4307 &  0.3966 & -0.2336 & -0.2708 & -0.2031 &    -   &    -   &  -   \\
     & 0.004 & 0.05 & 0.06 & 0.06 &  0.004  &  0.039  &  0.061  &  0.0112 &  0.0119 &  0.0226 &        &        &      \\
7562 & 0.304 & 3.09 & 2.71 & 1.74 & -0.0125 & -0.0607 & -0.0134 & -0.0400 & -0.0668 &  0.0390 &    -   &    -   &  -   \\
     & 0.003 & 0.04 & 0.04 & 0.04 &  0.003  &  0.029  &  0.041  &  0.0083 &  0.0083 &  0.0153 &        &        &      \\
7619 & 0.344 & 3.18 & 3.19 & 1.41 &  0.0275 &  0.2243 & -0.3434 & -0.0002 &  0.0146 &  0.2799 & -22.35 & -23.36 & 0.00 \\
     & 0.003 & 0.04 & 0.05 & 0.03 &  0.003  &  0.033  &  0.031  &  0.0086 &  0.0073 &  0.0118 &        &        &      \\
7626 & 0.350 & 2.99 & 2.91 & 1.49 &  0.0335 & -0.0107 & -0.2634 &  0.1535 &  0.0013 &  0.1914 & -22.36 & -23.36 & 2.60 \\
     & 0.003 & 0.03 & 0.04 & 0.04 &  0.003  &  0.026  &  0.041  &  0.0080 &  0.0082 &  0.0153 &        &        &      \\
7785 & 0.307 & 3.02 & 2.95 & 1.63 & -0.0095 &  0.0243 & -0.1234 & -0.0774 & -0.0785 &  0.1402 & -22.22 & -23.18 &  -   \\
     & 0.003 & 0.04 & 0.05 & 0.05 &  0.003  &  0.033  &  0.051  &  0.0088 &  0.0097 &  0.0189 &        &        &      \\
 & & & & & & & & & & & & &\\
\hline
\hline
\end{tabular*}
\end{center}
\normalsize
\label{tab6}
\end{table*}

If this age limit is real and not caused by statistical and/or selection
effects due to the scarce population of galaxies under consideration,
does it trace the locus of non-interacting galaxies, along which we
see the pristine star formation? If so, recalling the meaning of the age
parameter in usage here this implies that star formation either started later
or continued longer at decreasing galaxy mass (cf. Bressan et al. 1996 for a
similar suggestion).

All other galaxies are much scattered along the age axis. Indeed most of them
have sign of interaction or rejuvenation (Sweitzer \& Seitzer 1992).  Does it
imply that more recent star formation has occurred altering \Hbeta\
and age assignment in turn ?

M32 is an ambiguous case because either it could represent the continuation of
the trend shown by the old galaxies, in which star formation lasted till a
recent past or it could have been rejuvenated by a recent episode. As compared to
NGC~4649 there is a factor 4.0 in between. Assuming the canonical age of 15 Gyr
for the oldest galaxies, M32 terminated its star formation history or suffered
from star formation about 3.75 Gyr ago. The possibility that M32 contains a 
significantly younger stellar component is also indicated by studies of the 
classical color-magnitude diagram (CMD) of the resolved stars.
Although care must be paid because those CMDs do not refer to the central
region of M32,  an age of $4\div 5$ Gyr was estimated by Freedman (1989) and
Freedman (1992). See also Elston \& Silva (1992), Davidge \& Jones (1992), 
and Hardy et al. (1994) for similar conclusions.
Based on the population synthesis technique and the mean colors of M32,
O'Connell (1988 and references therein) suggested an age of about 6 Gyr.
Along the same vein Bressan et al. (1994) argued that the bulk of stars 
have ages as old as those typical of globular clusters say in the range 
$13\div 15$ Gyr, with a younger component which cannot be older than 5 Gyr 
and younger than 1 Gyr. This latter boundary is set by the UV properties 
of M32 which has (1550-V)=4.5 (Burstein et al. 1988). 
Remarkably, this estimate is consistent with the recent study 
by Grillmair et al. (1996) (of a field in M32 taken at the distance of about
$2\times R_e$), who find  that the CMD of this region  is consistent with a 
luminosity weighted age of about 8.5 Gyr and [Fe/H]=-0.25, however with 
some evidence for another component with [Fe/H]=0 for which younger ages 
cannot be excluded. As far as the very central part of M32 is concerned, 
an independent estimate of the age is still missing to our knowledge.
\littleskip

{\it Metallicity-magnitude plane: Z-$M_V$}. The relationship between 
$\rm \Delta \log(Z/Z_{\odot})$ and $\rm M_V$ is shown in Fig.~\ref{zeta_mv}.
According to the best fit of the data (thick dashed line)
the metallicity seems to weakly increase with the luminosity (mass) of the 
galaxy. Indeed, for the sample in usage, the very existence of this 
correlation depends on the inclusion of M32. Therefore, we will not 
insist on it.

Limiting the inspection to the group of galaxies that where used to argue about 
the  age limit in Fig.~\ref{age_mv}, all but NGC~4478 (and M32) are more metal-rich 
than the mean value of the sample but showing no particular trend with the 
luminosity. 

Worth noticing  is the case of all galaxies brighter than
$\rm M_V=-20$ and apparently younger than the mean age (cf. Fig.~\ref{age_mv})
whose metallicity is in contrast  above the mean value (but for NGC~3377 whose 
$\rm \Delta \log(Z/Z_{\odot})\simeq 0$). The remaining objects scatter 
above and below the mean value. Does this suggest that galaxies suffering 
from subsequent episodes of star formation have further increased their 
metallicity ? Furthermore, we call attention on  NGC~507, the brightest galaxy 
in the sample, which has age and metallicity only slightly below and above the 
mean, respectively. 

Going back to the apparent lack of a positive correlation between metallicity 
and luminosity, if we consider only those objects of the age group that are
also members of the Virgo cluster and therefore distance and absolute magnitudes
are less of a problem (namely NGC~4374, NGC~4478, and NGC~4552) they seem to be
more metal-rich at increasing luminosity.

Does the above inspection imply that the canonical trend ``metallicity increasing 
with luminosity'' holds only for quiescent galaxies (provided that their
absolute luminosity is well determined) ? And in all other cases the relation is
blurred  by recent stellar activity and/or uncertainties in their absolute
magnitudes ? More data are required to answer this question. 
\littleskip

{\it Enhancement-magnitude plane: [Mg/Fe]-$M_V$}. Fig.~\ref{mgfe_mv} shows the 
relation between $\rm \Delta [Mg/Fe]$ and $\rm M_V$. We start noticing that 
with the exception of NGC~4478 (and M32), all other galaxies of the group 
defining the age limit in Fig.~\ref{age_mv} are more enhanced in \alfa\ than 
the mean value of the sample, whereas the remaining  galaxies 
scatter below and above it. It appears that not necessarily galaxies with 
high metallicity are also enhanced in \alfa, even if from the results shown 
in Fig.~\ref{log_z_mgfe} some trend of this kind should hold on a broad sense. 
Furthermore, galaxies whose nuclear age is below the mean value not necessarily 
are more enhanced in \alfa. This can perhaps  be  explained recalling that star 
formation over periods of time longer than about 1 Gyr easily wipes out the signature 
type II supernov\ae\ in the abundance ratio $\rm [\alpha/Fe]$.
As far as the expectation that [Mg/Fe]  increases with the galaxy luminosity
is concerned, no such relation is noticed at a first sight. However, 
restricting ourselves only to galaxies members of the age limit group and of 
the Virgo cluster (namely NGC~4374, NGC~4478, and NGC~4552), the above trend 
is recovered. The same considerations made for the Z-$M_V$ plane apply also 
to this case.

\section{ Towards absolute ranking: an attempt }
Having established the relative differences \Dlogt, $\rm \Delta \log(Z)$, 
and $\rm \Delta [Mg/Fe]$ from galaxy to galaxy, the absolute ranking
of the central regions of different galaxies can be attempted. 
The main problem is with the zero point of the age, metallicity and 
enhancement scale. We take M32 as the reference object. As already mentioned 
several observational hints concur to suggest that M32 likely suffered from an 
episode or continued to form stars as recently as $4\div 5$  Gyr ago
(Freedman 1992). As far as the metallicity is concerned, it seems to  span a 
wide range. The lower limit in Freedman (1989) is [M/H]=-0.5 with an upper 
limit up to [M/H]=0.1 or even higher than that according to Bica et al. (1991). 
There is no direct indication of [Mg/Fe] (at least to our knowledge). For the 
purposes of the present discussion, we adopt as zero points: 4 Gyr for the age,
$\rm [Mg/Fe]_{M32}=0$ for the enhancement in \alfa, and $\rm [Fe/H]_{M32}=-0.25$ 
for the iron content or with the aid of equation (1), in which the term 
$\rm (X/X_{\odot})$ is neglected and $\rm Z_{\odot}=0.016$, $\rm Z_{M32}=0.0128$.  
The absolute values assigned to the age, metallicity, and [Mg/Fe]
of all other galaxies in the sample can be easily re-scaled if the zero 
point is going to change.

The age is assigned  with the aid of the grids of  SSP  drawn in Fig.~\ref{age_mv}.
The  metallicity and [Mg/Fe] are assigned by simply shifting the vertical 
axis in Figs.~\ref{zeta_mv} and \ref{mgfe_mv}. The results are summarized in Table~4
limited to the group of galaxies that in Fig.~\ref{age_mv} define the  age
boundary. 

\begin{table}
\begin{center}
\caption{Estimated ages, metallicities, and enhancements of \alfa
         for the group of galaxies defining the old age edge in Fig.~17 }
\begin{tabular}{l| c r r  }
\hline
\hline
 & & & \\
\multicolumn{1}{l|}{NGC} &
\multicolumn{1}{l}{Age} &
\multicolumn{1}{c}{$<Z>$} &
\multicolumn{1}{c}{[Mg/Fe] }\\ 
 & & & \\
\hline
 & & & \\
221   &    4.0   &  0.0128 &  0.00   \\
2778  &    8.4   &  0.0216 &  0.62   \\
3379  &   11.4   &  0.0180 &  0.51   \\
3608  &   10.0   &  0.0184 &  0.49   \\
3818  &   10.0   &  0.0184 &  0.46   \\
4261  &   17.4   &  0.0193 &  0.45   \\
4374  &   14.0   &  0.0245 &  0.49   \\
4478  &    9.0   &  0.0150 &  0.28   \\
4552  &   12.9   &  0.0219 &  0.57   \\
4649  &   17.3   &  0.0224 &  0.58   \\
7619  &   18.8   &  0.0181 &  0.42   \\
 & & & \\
\hline
\hline
\end{tabular}
\end{center}
\label{tab7}
\end{table}

\section { Isolation or mergers? }
Perhaps the most intriguing question to address about galaxies is
which of the two main avenues for their formation and evolution has prevailed
as the dominant mechanism, i.e. isolation or mergers.
The fading lines in Fig.~\ref{age_mv} help to visualize in the age-magnitude
diagram the path followed by an evolving galaxy under a number of different 
scenarios.

If a galaxy forms and evolves in isolation, and suffers from a number of
episodes of star formation (from one to several or even continuous) it would
simply slide up and down along its fading line  according to the stage of
stellar activity at which the galaxy is detected with little transversal shift
caused by chemical enrichment. Complications due to the possible 
presence of gas can be neglected here. Once star formation is over, the 
recovery position in the age-luminosity plane depends on the age and amount 
of stars formed in each episode. As already said the data are compatible with
this scheme suggesting that there are two groups of galaxies: those with no
sign of rejuvenation in which either the overall duration increases or the 
epoch of the last episode of star formation gets closer to the present at 
decreasing galaxy mass, and those with signs of rejuvenation which are obviously
scattered in this diagram.

The case of hierarchical merging is more difficult to discuss because the path
is determined by the mass and evolutionary stage of the merging galaxies, and
the amount of star formation taking place during  the merger event. Let us
suppose for the sake of simplicity that two identical units (same age, same
composition, and say $10^{10} \times M_{\odot}$ mass each) merge  triggering
star formation in the composite system. The resulting galaxy will brighten 
because of the increased mass and ongoing star formation and slightly
redden because of the increased metallicity. As soon as star formation is over, 
on a short time scale (from a few $10^8$ yr  to 1 Gyr at most) the new galaxy will fade 
back to a new position which is characterized by a younger age (it depends on
the amount of newly formed stars) and a brigheter magnitude.
In principle there is no contradiction between this scenario and the bunch of 
galaxies scattered in this diagram toward much younger ages. The problem is with 
their recovery position if this latter corresponds to the group of galaxies 
with no sign of rejuvenation. Indeed if the seed galaxies had the same age, we 
would expect the daughter galaxies to cluster in the age-luminosity plane 
along a nearly horizontal line (or mildly inclined toward younger ages). 
In contrast, it seems as if bright, high mass galaxies had their
merger adventure long ago, whereas the faint, low mass objects did it in a
more recent past, but starting from seed galaxies with small stellar content.
This is equivalently to say that the bulk of star forming activity took place
more and more toward the present at decreasing mass of the galaxy.

Furthermore, {\it why don't we see in this diagram any low mass galaxy (the
potential seeds of a bigger galaxy in the hierarchical scheme) at the same age
range of the big galaxies ? Is this only due to selection effects in our
sample  or other  causes are to be considered? } Nowadays answering this question 
is difficult owing to the lack of sufficient data with precise measurements of \Hbeta,
\Mg2 and \MFe.

\section{ SN-driven winds or variable IMF? }
Long ago Larson (1974) suggested that the color-magnitude relation (CMR) of early 
type galaxies (cf. Bower et al. 1992 for  Virgo and Coma galaxies), is the 
consequence of SN-driven galactic winds. In the classical scenario, isolation 
and constant IMF, massive galaxies eject their gaseous content much later and 
get higher mean metallicities than the low mass ones. The implication is that 
the global duration of the star forming activity increases with the  galaxy mass. 

The alternative of the CMR being an age sequence in the sense that blue galaxies 
have started star formation much later than in red ones has be proved to disagree 
with the redshift evolution of early type galaxies in the HST Deep Field (cf. 
Kodama \& Arimoto 1996). 

Bright ellipticals seem to be more enhanced in \alfa\ than the faint ones 
($\alpha$-enhancement). The classical explanation of it (cf. Matteucci
1997 for a recent review of the subject) is based on the classical view of star 
formation in galaxies, i.e. constant IMF and time dependent star formation rate,   
the different time scale with which the galactic gas is contaminated by the products 
of the short-lived Type II supernovae (main producer of \alfa) and the long-lived 
Type I supernovae (main producers of Fe), that the overall duration of the star 
forming activity is short in massive ellipticals and long in the low-mass ones. 

The present analysis seems to lead to a more complex scenario in which the 
trends implied by the CMR and enhancement in \alfa\ are perhaps 
simultaneously recovered  only for {\it quiescent}  galaxies  (provided their 
magnitudes are well determined), whereas in all other cases, in which \Hbeta\ 
and $\Sigma$ indicate signatures of recent stellar activity, rejuvenation 
and/or interactions, the above key relations are blurred by side effects hard 
to quantify. The results of our study somehow weaken the classical SN-driven wind 
model based on monolithic star formation in isolation (and constant IMF), but 
at the same time are not fully compatible with the  hierarchical merger mechanism.  

Seeking for a coherent explanation of the various observational hints, we propose
the following scheme. All galaxies have begun to form stars at the same time but,
depending on their mass, the process has continued (maybe in discrete
episodes) over different periods of time. Massive galaxies did it in the far
past and the star forming period ceased very soon.  Low mass galaxies
started at the same epoch, but continued for longer periods of time (the
duration increasing at decreasing galactic mass). Subsequently, galaxies of any 
luminosity (mass) may have undergone additional episodes of star formation, 
depending on circumstances that cannot be singled out by this kind of analysis, 
which reflect themselves in the large dispersion in the age-$M_V$ plane.
As far as the metallicity and enhancement in \alfa\ are concerned, although the
tendency to increase with the galaxy luminosity (mass) can be noticed in the case of
inactive galaxies, as required by the CMR $\alpha$-enhancement problem, it seems 
as if each galaxy had its own individual history of enrichment in heavy and \alfa. 

In regard to this, the attempt by  Chiosi et al. (1998) to reconcile the 
SN-driven wind model with the pattern of abundances and the tilt of the
Fundamental Plane of galaxies is worth mentioning. They supposed that
the IMF instead of being constant can vary with the galxy environment, and 
presented new models of elliptical galaxies based on the IMF  by 
Padoan et al. (1997), which depends on the density, temperature and 
velocity dispersion of the medium in which stars are formed. 
In brief,  in a hot, rarefied medium the Padoan et al. (1997) IMF is 
more skewed towards the high mass end than in a cool, dense medium. This kind 
of situation is  met passing from a high mass (low mean density) to  a low mass 
(high mean density) galaxy or from the center to the periphery of a given galaxy. 
The models based on this IMF can account for a number of observed chemical and photometric 
constraints for elliptical galaxies. In fact, they predict the onset of galactic 
winds and consequent termination of the star forming period  earlier in massive 
galaxies than in the low mass ones. The reason of it resides in the skewness of 
the IMF toward the high mass end that changes  with the mean gas density (galactic
mass and/or position within a galaxy) thus favoring in massive galaxies the
relative percentage of SN explosions and consequent heating of the gas to the
escape velocity. In this scheme massive galaxies despite their short duration
of stellar activity yet reach high metallicities and  [Mg/Fe] ratios. The
opposite occurs in the low mass ones. Unfortunately, Scalo et al. (1997) have
strongly questioned the Padoan et al. (1997) IMF, thus weakening the foundations 
of the Chiosi et al. (1998) scenario. However, according to Scalo et al. (1997) 
other model IMF could be constructed to give similar dependences. Whether they
would also lead to model galaxies like those by Chiosi et al. (1998)
cannot be said. All the problems are still there!

\section{Conclusions}

In this study we have  investigated the ability of the $\rm H_{\beta}$, \Mg2\ 
and \MFe\ diagnostics to assess the metallicity, [Mg/Fe] ratios, and ages of 
elliptical galaxies. First, we have tried to interpret the gradients in \Hbeta, 
\Mg2\,and \MFe\ across individual  galaxies. Second, we have tackled the problem 
of the  information hidden in the different values of \Hbeta, \Mg2, and \MFe\ 
observed in the nuclear regions of elliptical galaxies. The results of this study 
can be summarized as follows:

\begin{enumerate}
\item{ We provide basic calibrations for the variations $\rm \delta H_{\beta}$, 
$\rm \delta Mg_2$ and \dfem\ as a function of variation in age \Dlogt, 
metallicity $\rm \Delta \log(Z/Z_{\odot})$,  and $\rm \Delta [Mg/Fe]$ of SSPs 
whose application is of general use.}
\littleskip

\item{Limited to a small sample of objects for which the observational data 
are available, we analyze from a quantitative point of view how the
difference $\rm \delta H_{\beta}$, \dmg, and $\rm \delta \langle Fe
\rangle$ between the external and central values of each index 
translates into $\rm \Delta [Mg/Fe]$, $\rm \Delta \log(Z/Z_{\odot})$, and 
$\rm \Delta \log(t)$.  We find that out of six galaxies under examination, 
four have the nuclear region more metal-rich, more enhanced in \alfa, and 
younger (i.e. containing a significant fraction of stars of relatively young 
age) than the external regions. In contrast the remaining two galaxies have 
the nuclear region more metal-rich, more enhanced in \alfa\ but marginally 
older than the external zones. Whether this dichotomy in the age difference 
between the central and the external regions is a rule is difficult to assess 
at the present time owing to the very small number of objects for which the 
analysis has been feasible. Such a possibility has been envisaged by Bressan et 
al. (1996). We intend to reconsider the whole problem in a forthcoming paper 
utilizing the 20 galaxies of the Gonz\'ales (1993) catalog  for which gradients 
are available.}
\littleskip

\item{ The above calibration is used to explore the variation from galaxy to
galaxy of the nuclear values of \Hbeta, \Mg2, and \MFe\
limited to a sub-sample of the Gonz\'ales (1993) list. The differences \dhb,
\dmg, and $\rm \delta \langle Fe \rangle$ are converted into the differences
\Dlogt, $\rm \Delta \log(Z/Z_{\odot})$, and $\rm \Delta [Mg/Fe]$. Various
correlations among the age, metallicity, and enhancement variations are
explored. In particular we thoroughly examine the relationships 
$\rm \Delta\log(t)-M_V$, $\rm \Delta \log(Z/Z_{\odot})-M_V$, and $\rm \Delta
[Mg/Fe]-M_V$. We advance the suggestion that a sort of age limit is likely to exist
in the  $\rm \Delta\log(t)-M_V$ plane, traced by galaxies with mild or no 
sign of rejuvenation. In these objects, the duration of the star
forming 
activity is likely to have increased at decreasing galactic mass. Limited to
these galaxies and provided the luminosity is well determined, 
the mass-metallicity sequence implied by the CMR is recovered, 
likewise for the $\alpha$-enhancement-luminosity relation suggested by
the  gradients in \Mg2\ and \MFe. For the remaining galaxies the situation 
is more intrigued: sporadic episodes of star formations are likely to
have occurred  scattering the galaxies in the space of age,
metallicity,  and  [Mg/Fe]. 
Owing to the very small group of galaxies from which the suggestion for the
existence of an age limit is drawn,
we consider it as provisional. Work is in progress to check it on the larger
sample of data by Trager (1998), in which smaller elliptical galaxies than in 
the Gonz\'ales (1993) sample are present to our disposal.}

\item{ 
The results of this study are
discussed in regard to  predictions from the merger and isolation models of
galaxy formation and evolution highlighting points of difficulty with each
scheme. Finally, the suggestion is advanced that models with an IMF
that favors higher mass stars in massive elliptical galaxies, and lower
mass stars in low-mass ellipticals, might be able to alleviate some of the
difficulties encountered by the standard SN-driven galactic wind model and
lead to a coherent interpretation of the data. }
\end{enumerate}

\acknowledgements{
We deeply thank Dr. Guy Worthey for his constructive referee report. His
remarks greatly improved the original version of the paper.
C.C.\ is pleased to acknowledge the hospitality and stimulating environment 
provided by ESO in Garching where this paper was finished during sabbatical 
leave from the Astronomy Department of the Padua University. This study has
been financed by the Italian Ministry of University, Scientific Research and 
Technology (MURST), the Italian Space Agency (ASI), and the TMR
grant ERBFMRX-CT96-0086 from the European Community.}

\newpage
\newpage

\end{document}